\documentclass[prd,a4paper,superscriptaddress,aps,showpacs,twocolumn,floats,nofootinbib]{revtex4}
\usepackage{epsfig}
\usepackage{graphicx}
\usepackage{amsmath}
\usepackage{amssymb}
\usepackage{color}
\usepackage{soul}
\usepackage{hyperref}
\usepackage{url}
\usepackage{float}

\begin{document}

\title{Impact of Redshift Space Distortions on Persistent Homology of cosmic matter density field}

\author{Fatemeh Abedi}

\affiliation{Department of Physics, Shahid Beheshti University,  1983969411, Tehran, Iran}

\author{Mohammad Hossein Jalali Kanafi}

\affiliation{Department of Physics, Shahid Beheshti University,  1983969411, Tehran, Iran}

\author{S.M.S. Movahed}
\email{m.s.movahed@ipm.ir}

\affiliation{Department of Physics, Shahid Beheshti University,  1983969411, Tehran, Iran}
\affiliation{School of Astronomy, Institute for Research in Fundamental Sciences (IPM), P. O. Box 19395-5531, Tehran, Iran}
\affiliation{Department of Mathematics and Statistics, The University of Lahore,
    1-KM Defence Road Lahore-54000, Pakistan}

\date{\today}

\begin{abstract}

By employing summary statistics obtained from Persistent Homology (PH), we investigate the influence of Redshift Space Distortions (RSD) on the topology of excursion sets formed through the super-level filtration method applied to three-dimensional matter density fields. The synthetic fields simulated  by the Quijote suite in both real and redshift spaces are smoothed by accounting for the Gaussian smoothing function with different scales. The RSD leads a tendency for clusters ($\tilde{\beta}_0$) to shift towards higher thresholds, while filament loops ($\tilde{\beta}_1$) and cosmic voids ($\tilde{\beta}_2$) migrate towards lower thresholds. Notably, $\tilde{\beta}_2$ exhibits greater sensitivity to RSD compared to clusters and independent loops. As the smoothing scales increase, the amplitude of the reduced Betti number curve ($\tilde{\beta}_k$) decreases, and the corresponding peak position shifts towards the mean threshold. Conversely, the amplitude of $\tilde{\beta}_k$ remains almost unchanged with variations in redshift for $z\in[0-3]$. The analysis of persistent entropy and the overall abundance of $k$-holes indicates that the linear Kaiser effect plays a significant role compared to the non-linear effect for $R \gtrsim 30$ Mpc $h^{-1}$ at $z=0$,  whereas persistent entropy proves to be a reliable measure against non-linear influences.

\end{abstract}

\maketitle

\vspace{0.1cm}
The presence of high-precision data, especially in multi-dimensional formats, presents inherent challenges in maximizing the extraction of information. A structured framework for handling such data entails the formulation of an adequate feature vector or summary statistics, which is then complemented by the implementation of either data-based or theory-based modeling methodologies. A comprehensive evaluation of the formation of large-scale structures (LSS) \cite{peebles2020large,Bernardeau:2001qr} inherently relies on the principles of stochastic fields ($\mathcal{F}$) across multiple dimensions\footnote{A typical $(d+D)$-Dimensional stochastic field which is represented by $\mathcal{F}^{(d,D)}$, is a measurable mapping from probability space into a $\sigma$-algebra of $\mathbb{R}^{d}$-valued function on $\mathbb{R}^D$ Euclidean space \cite{adler81,adler2011topological,adler2010persistent}. The indices  $d$ and $D$ are related to the $d$-dependent parameters and $D$-independent parameters, respectively.}.

The decomposition of different categories of measurable components within LSS, expressed through functions that comprise an orthogonal and complete set, presents both benefits and drawbacks. When employing Fourier or Spherical Harmonics basis functions, the localized nature of the field tends to lose clarity \cite{2024arXiv240313985Y}. 
Furthermore, while conventional one-point statistics can yield significant insights, they are inherently limited in their ability to capture the complexities that are fundamentally represented in higher-order statistics. The unweighted Two-Point Correlation Function (TPCF) serves as a fundamental extension for understanding the characteristics of pairs of features \cite{peeb80,kaiser1984spatial,Bardeen:1985tr} and it has well-defined relation to the weighted TPCF \cite{rice1954selected,szalay88,desjacques2018large}.
(see also \cite[and references therein]{2021MNRAS.503..815V}). The notion of excursion and critical sets has long-standing to quantify the geometrical aspects of cosmological fields \cite{matsubara2003statistics,Pogosyan:2008jb,Codis:2013exa,matsubara2020statistics,2021MNRAS.503..815V,2024MNRAS.528.1604S}. Among the early measures introduced to characterize the morphology of stochastic fields are crossing statistics, which include up-crossings, down-crossings, conditional crossings, and contour crossings \cite{rice44a,rice44b,Bardeen:1985tr,Bond:1987ub,ryden1988,ryd89,matsubara2003statistics}.
The corresponding size, shape, connectedness, and boundaries encompass by the un-weighted and weighted morphologies \cite{2024ApJ...963...31K,2023arXiv231113520J}.

Considering the principles of motion invariance, additivity, and
conditional continuity as articulated in {\it Hadwiger's theorem}, the
$(1+D)$ functionals, referred to as Minkowski Functionals (MFs), are
employed to quantify the morphological characteristics of a
$D$-dimensional field \cite{mecke1993robust,schmalzing1997beyond}.
Minkowski Valuations (MVs) are presented through the relaxation of
particular restrictions inherent in {\it Hadwiger's theorem}
\cite{McMullen1997,Beisbart2002}. In more recent studies, Minkowski
Tensors (MTs) as an extension of Minkowski Functionals (MFs) and a
weighted form of Minkowski Variables (MVs), named as the {\it
conditional moments of first derivative},  have found applications
in multiple areas of cosmology \cite[and references
therein]{matsubara1996genus,codis2013non,appleby2018minkowski,appleby2019ensemble,Appleby2023,2024ApJ...963...31K}.

In the context of high precision observational cosmology, the ongoing and future surveys namely DESI\footnote{\texttt{http://www.desi.lbl.gov}}, PFS\footnote{\texttt{http://pfs.ipmu.jp}}, the Roman Space Telescope\footnote{\texttt{http://wfirst.gsfc.nasa.gov}}, Euclid\footnote{\texttt{http://sci.esa.int/euclid}}, and CSST\footnote{\texttt{http://nao.cas.cn/csst}} \cite{2011SSPMA,2019ApJ883} offer a significant opportunity for a thorough examination of galaxy clustering. These efforts are expected to yield profound insights into the processes of structure formation in the universe, thereby linking the initial primordial conditions to the later epochs of cosmic evolution.  Also the joint analysis with the next generation of observations such as CMB Stage IV \cite{2019arXiv190704473A,2017arXiv170602464A}, has the potential to assess the extension of the vanilla-$\Lambda$CDM model.

Over the past few years, the notable progress in achieving high levels of accuracy has resulted in the emergence of various anomalies, tensions, and degeneracies. Consequently, scientists have been motivated to develop and execute comprehensive measures to alleviate these issues \cite{2022NewAR..9501659P,2023CQGra..40i4001K}. The motivation to develop and employ sophisticated methodologies arises from the necessity to differentiate among various theories, impose stringent constraints on parameters, and mitigate or at least lessen the challenges posed by degeneracies and tracers.

The application of geometrical and topological measures in the analysis of cosmological stochastic fields presents a promising approach to reconciling what we observed in various surveys with the predictions derived from theoretical models \cite{kaiser1984spatial,Bardeen:1985tr,Bernardeau:2001qr,matsubara2003statistics,matsubara2020statistics,matsubara1996genus,codis2013non,appleby2018minkowski,appleby2019ensemble,Appleby2023,2024ApJ...963...31K}. In redshift surveys, the positions of cosmic structures are determined based on redshift values rather than direct distance measurements. Consequently, the actual locations of galaxies differ from their observed positions, particularly in the regions influenced by peculiar velocities. The phenomenon previously mentioned is referred to as Redshift Space Distortions (RSD), which can be classified into two categories: the large-scale effect known as the ``linear Kaiser effect" \cite{Kaiser1987} and the small-scale effect termed the ``Finger of God (FoG) effect" \cite{Jackson1972, peebles2020large}. The Kaiser effect refers to the elongated configuration of clusters observed along the line-of-sight, while the Finger of God (FoG) effect is predominantly caused by the random movements of galaxies within virialized clusters at smaller scales  \cite{sargent1977statistical,hamilton1998linear}. At first glance, it is probably figured out that RSD increases the ambiguities of extracting information encoded in the exploring of the LSS but as long as utilizing suitable tools for examination, one can  extract statistical information to constrain corresponding cosmological parameters \cite{hamilton1998linear,Bernardeau:2001qr}.

In order to elucidate the importance of RSD for different cosmological interpretations, it is useful to highlight a few recent studies, including: Examining the linear growth rate of density fluctuations by calculating the correlation between the redshift distortions and cosmic mass distribution \cite{hamaus2022euclid,panotopoulos2021growth};
Improving the observational constraints on the relevant cosmological parameters by combining the weak lensing and baryon acoustic oscillations  with RSD \cite{eriksen2015combining}; Probing the cosmic expansion considering  the Alcock-Paczynski effect and RSD  \cite{Song2015}; Assessing the primordial non-Gaussianity through the  RSD \cite{tellarini2016galaxy,bharadwaj2020quantifying}; Computing the 2D redshift-space galaxy power spectrum and separating non-linear growth and RSD effects \cite{2016MNRAS.457.1076J}; Quantifying the redshift space distortions of the bispectrum \cite{bharadwaj2020quantifying,mazumdar2020quantifying,mazumdar2023quantifying}.
The presence of non-Gaussianity and anisotropy resulting from the distortions of redshift space, particularly when nonlinear effects become significant within structures, serves as a compelling reason to incorporate features that remain unaffected by RSD. Consequently, it is possible to differentiate between physical phenomena that arise from early processes and those that occur at later times. The statistical analysis of isodensity contours in redshift space has demonstrated that their shapes remain consistent with those observed in real space, with the only variation being in the amplitudes influenced by  RSD \cite{matsubara1996statistics}.  The MTs have been utilized to capture the morphology of LSS in both real and redshift spaces \cite{appleby2018minkowski, appleby2019ensemble,Appleby2023,ApplebySDSS}.  The MFs of LSS on both (quasi-)linear and non-linear scales due to the peculiar velocities has been computed in  \cite{2022PhRvD.105j3028J}.
\begin{figure*}
    \begin{center}
        \includegraphics[scale=0.3]{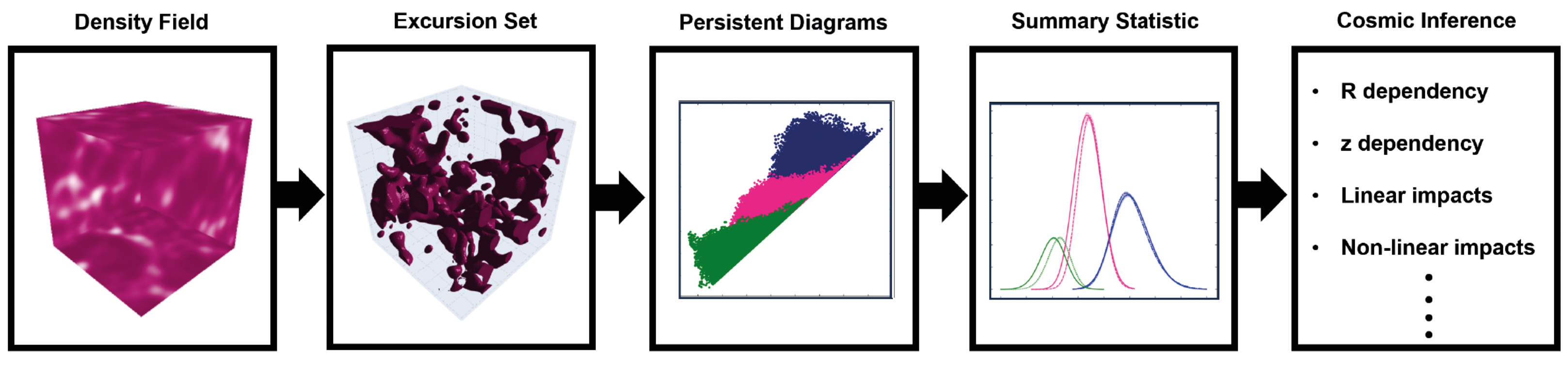}
        \caption{The pipeline considered in this paper.  From the left to right: the matter density fields are created in both real and redshift spaces for different redshift epochs and they are smoothed by various values of smoothing scales; considering super-level filtration, the excursion set is constructed for different thresholds; persistence diagrams for connected regions, closed loops and closed surfaces are extracted; The proper summary statistics according to assumed  PH vectorization is computed; finally a diverse cosmological inferences are performed.}
        \label{fig:pipeline}
    \end{center}
\end{figure*}

The convergence of algebraic topology and computational geometry presents a valuable opportunity to deepen our comprehension and yields profound insights into the quantitative evaluation of diverse cosmological domains \cite{2024arXiv240313985Y}. Topological characteristics of a manifold describe properties that are preserved under continuous deformations.
The cosmological mass distribution field has been analyzed by focusing on topological invariants such as the genus and the Euler characteristic \cite{1986ApJ...306..341G,1986ApJ...309....1H,1989ApJ...340..625G,pranav2019topology,2021MNRAS.507.2968W}.

Simplicial complexes provide a framework for associating shapes with
both discrete and continuous high-dimensional datasets, allowing for
the computation of various dimensional voids based on specified
proximity parameters (thresholds). This methodology  has gained
considerable attention because of its capacity to encapsulate proper
information \cite[and references
therein]{2018JCAP...03..025C,2021MNRAS.507.2968W,2021A&A...648A..74H,2022A&A...667A.125H,2023MNRAS.522.2697T,2022JCAP...10..002B}.
Persistent Homology (PH), a technique adopted in Topological Data
Analysis (TDA)
\cite{edelsbrunner2022computational,zomorodian2005topology11,wasserman2018topological},
and it provides an advanced framework for analyzing the complex data
sets by utilizing topological invariants. This approach facilitates
the extraction of significant information, thereby various
characteristics of the underlying structures can be computed. PH
serves as a mathematical instrument for the analysis and
comprehension of LSS by identifying and monitoring the persistence
of topological features, including clusters, filaments, and voids,
across various scales
\cite{van2011,Xu2019,2021MNRAS.507.2968W,biagetti2021persistence,2023MNRAS.522.2697T,2023MNRAS.520.2709E}.
The application of PH to distinguish between models of hot and cold
dark matter has been investigated by \cite{Cisewski-Kehe2022PhRvD}.
Additionally, PH has been employed to investigate non-Gaussianity
\cite{feldbrugge2019stochastic,biagetti2021persistence,2022JCAP...10..002B}.
In addition to topological measures, noteworthy geometrical criteria
have been utilized in cosmological analysis, especially to impose
stringent constraints on cosmological parameters, such as the total
mass of neutrinos \cite{2023PhRvX..13a1038P}. Recent studies have
highlighted the multiscale topological features of large-scale
structures (LSS) and have employed joint analysis of power spectrum
and bispectrum statistics to refine cosmological parameter estimates
\cite{2024arXiv240313985Y}.

Numerous studies focus on the application of topological data
analysis across a wide array of disciplines, including a
comprehensive review article that highlights these efforts
\cite{2018AnRSA...5..501W} and particularly the TDA focusing on the
cosmological applications \cite{2015PhDT.......250P}. The TDA has
been used to examine the spatial distribution of galaxies
\cite{2020arXiv200602905K}. The Significant Cosmic Holes in Universe
has been carried out to recognize the cosmic void and filaments via
TDA  \cite{Xu2019}.  Using the multi-scale topology of large-scale
structures,  the primordial non-Gaussianity of the local and
equilateral type has been explored \cite{2022JCAP...10..002B}. The
formalism, evolution and classification of Persistent topology of
the reionization bubble network have been done
\cite{2019MNRAS.486.1523E,2023MNRAS.520.2709E} the evolution of
so-called connectivity and the topology of structure of the cosmic
has been checked  \cite{2021MNRAS.507.2968W}. The persistent
homology of LSS  in the SDSS fields was investigated
\cite{kimura2017quantification}. The coherent multiscale
identification of all types of cosmological and astrophysical
structures has been done through persistent cosmic web
\cite{2013ascl.soft02015S,2013arXiv1302.6221S,2011MNRAS.414..350S,2011MNRAS.414..384S}.
The Betti numbers of cosmic shear cosmological analysis with Euclid
has been done  \cite{2023A&A...675A.120E}. The connectivity of the
cosmic web by using topology has determined
\cite{2017MNRAS.465.4281P, 2021arXiv210908721P}

In examining the RSD, it is important to emphasize that the theoretical framework for large-scale structure is predominantly constructed upon the characteristics observed in real space. Conversely, from an observational standpoint, the data obtained from redshift surveys reveals significant distortions in the observed field. To achieve a precise evaluation and to  compare theoretical predictions with observational data, it is imperative to thoroughly characterize the non-Gaussianity and anisotropy induced by redshift space distortions (RSD). This characterization is vital for distinguishing among primordial physics, foreground effects, and phenomena occurring at later epoch. In light of this significant disparity, a pertinent question emerges concerning the effects of RSD on the topological attributes of distinct components of the cosmic web, which includes the underlying field as well as various tracers. The modeling of linear RSD is significantly more straightforward than that of the non-linear component. Consequently, identifying criteria that are resilient to non-linear effects while remaining sensitive to the linear influence of Kaiser can yield valuable insights for cosmological interpretations. In this context, driven by the objective to assess the connected components (analogous to clusters), independent loops (analogous to filaments), and isolated closed surfaces (indicative of voids) within the matter density field in both real and redshift space, we employ a specific PH vectorization. This approach enables us to calculate the topological invariants alongside complementary measures within the algebraic-topological framework of the density matter field as simulated by the Quijote suite in both real and redshift domains.  The pipeline is conducted for a variety of redshift intervals and different smoothing scale values. The main novelties and advantages of our research are categorized as below:

(1) A comprehensive pipeline will be developed to ensure the input field is adequately prepared for the implementation of PH analysis, which will also encompass an exclusive summary statistics (Fig. \ref{fig:pipeline}).

(2) Furthermore, our analysis will include an assessment of the influence exerted by the linear and non-linear segments of RSD on the topological features of the matter density field, as simulated by the Quijote suite.

(3) We systematically assess the components of our summary statistics that are either affected or unaffected by the non-linear characteristics of RSD. In addition, we will provide clarity regarding the impact of smoothing scales and the redshift dependence inherent in our findings.

The subsequent sections of this paper are structured as follows: Section \ref{sec:rsd} provides a concise overview of RSD, along with the introduction of our notation for matter density fields in both redshift and real spaces. In Section \ref{sec:ph}, we examine the concept of homology through the lens of excursion set theory and its relevance to discrete cosmological datasets. Section \ref{sec:data} elucidates the mock data produced by N-body simulations. Section \ref{sec:imp} focuses on the PH of matter density fields in both real and redshift spaces. Finally, the summary and conclusions are presented in Section \ref{sec:con}

\section{Redshift Space Distortions}\label{sec:rsd}
\begin{figure*}
    \begin{center}
        \includegraphics[scale=0.3]{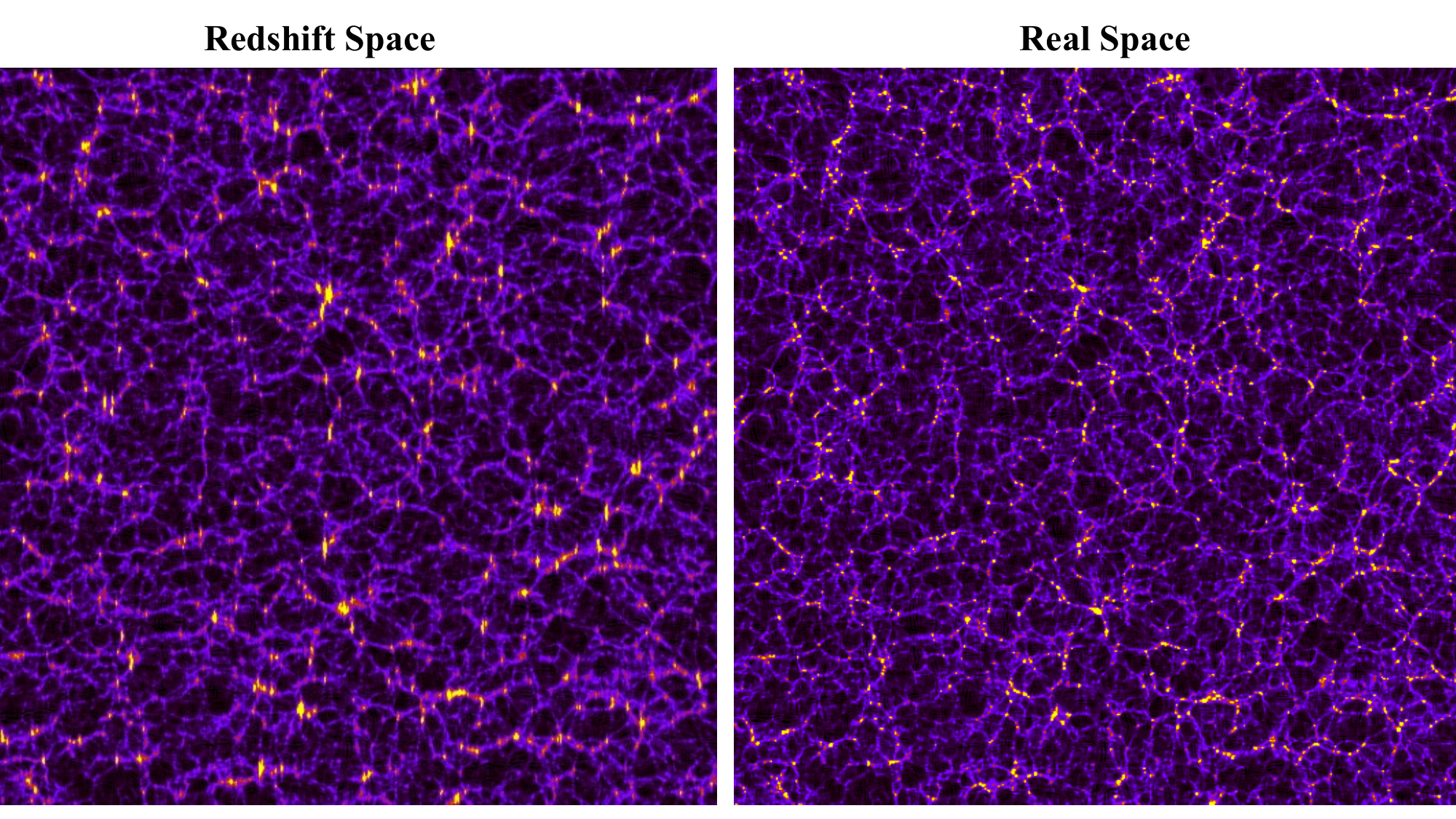}
    \end{center}
    \caption{The slice with thickness equates to $9.76$ Mpc $h^{-1}$ of a mock 3-dimensional matter density field with the $\Lambda$CDM cosmology. The redshift for this map is zero and therefore $\mathcal{B}=0.526$. The left panel shows the matter density field distorted by RSD phenomena, while the right panel is for the real space. The observer is located at the bottom of the figure. The box size is $1$ Gpc $h^{-1}$ with $512^3$ dark matter particles.}\label{fig:density}
\end{figure*}

Redshift surveys for LSS mainly provide two angular positions and redshift as an indicator of radial distance leading to have position as a function of redshift instead of  distance. The redshift is predominantly influenced by Hubble flow. The Hubble flow is also contaminated by peculiar velocity parallel to the line-of-sight. Consequently, the observed location of cosmic  structure compared to the real space position is modified via:
\begin{eqnarray}
\boldsymbol{s} = \boldsymbol{r} + \frac{\boldsymbol{v(\boldsymbol{r})}.\hat{\boldsymbol{n}}}{H} \hat{\boldsymbol{n}},
\label{eq:redshift position}
\end{eqnarray}
where $\hat{\boldsymbol{n}}$  is the line-of-sight direction, $H$ is the Hubble parameter and $\boldsymbol{v(\boldsymbol{r})}$ shows the peculiar velocity. Consequently the statistics of  corresponding structure such as matter density field is modified in the redshift space with respect to the real space. The Fourier transformation of  a typical tracer (density contrast) in the redshift space is expressed in terms of real space counterpart as follows:
\begin{eqnarray}
\tilde{\delta}^{(s)}(\boldsymbol{k}) = \tilde{O}_{s}(\mu,k\mu)\tilde{\delta}^{(r)}(\boldsymbol{k}).
\label{eq:anisotropic delta}
\end{eqnarray}
here  the operator $\tilde{O}_{s}$ is decomposed into the linear Kaiser part ($\tilde{O}_{\rm{lin}}$) and the non-linear part ($\tilde{O}_{\rm{nl}}$) in the plane-parallel approximation, as:
\begin{eqnarray}
\tilde{O}_{s}(\mu,k\mu)& =& \tilde{O}_{\rm {lin}}(\mu,k\mu)\times\tilde{O}_{\rm {nl}}(\mu,k\mu)\nonumber\\
&=&(1+\mathcal{B}\mu^2)\tilde{O}_{\rm {nl}}(\mu,k\mu).
\label{eq:operator}
\end{eqnarray}
where the redshift distortion parameter is $\mathcal{B}\equiv f/b$ and $f=d\ln(\delta)/d\ln a$ is the linear growth rate of the density contrast.  The bias factor ($b$) is unit for matter while $b\ne 1$ for  the biased tracers. The $\mu\equiv \hat{k}.\hat{n}$ is the cosine of the angle between the line-of-sight and the wave vector \cite{hamilton1998linear}. Eq. (\ref{eq:operator}) demonstrates that the density field in redshift space is distorted due to the linear and nonlinear terms. Our purpose is examining the PH vectorization of $\tilde{\delta}^{(s)}$ and  $\tilde{\delta}^{(r)}$ according to our pipeline illustrated by Fig. \ref{fig:pipeline}. To this end, we use the synthetic field generated by Quijote suite for different redshift snapshots smoothed by different smoothing scale values.
Fig.~(\ref{fig:density}) shows  the slices of a 3-dimensional matter density field at $z=0$ in redshift space (left) and its corresponding field in real space (right). The elongation along the line-of-sight known as FoG is obvious in the redshift space.

\section{The Foundation of Persistent Homology}\label{sec:ph}
In this section, we will explain the key aspects of Persistent Homology (PH)  for quantifying the topological invariants of discrete data sets,  according to our pipeline indicated by Fig. \ref{fig:pipeline}.  Considering a sample of matter particles on a cubic lattice including their corresponding locations at a given cosmic epoch, $t$ (or corresponding redshift, $z$). Our case study  consists of
dark matter particles extracted from N-body simulations divided into $N^3_{grid}$. We then allocate a number density contrast to each lattice cell ($\boldsymbol{r}$) for a given cosmic time ($t$) as follows:
\begin{equation}\label{eq:delta}
\delta^{\diamond}(\boldsymbol{r}, t) \equiv \frac{n^{\diamond}(\boldsymbol{r}, t) - \bar{n}^{\diamond}(t)}{\bar{n}^{\diamond}(t)}
\end{equation}
here $n^{\diamond}(\boldsymbol{r}, t)$ shows the number of particles at position $\boldsymbol{r}$ and given time $t$, while $\bar{n}^{\diamond}(t)$ is the mean value of particles number over the cubic lattice. Obviously, in such case $\delta^{\diamond} \in [-1, \infty)$. The $\diamond$ is replaced by $(s)$ and $(r)$ for redshift and real spaces.
The construction of continuous density field is accomplished by applying a convolution with a particular smoothing window function as:
\begin{equation}\label{eq:smoothed}
\delta_{\rm smoothed}^{\diamond}(\boldsymbol{r}, t)=\int d\boldsymbol{r'}\; \mathcal{W}(\boldsymbol{r}-\boldsymbol{r'};R)\;\delta^{\diamond}(\boldsymbol{r'}, t)
\end{equation}
where the $R$ is smoothing scale. We adopt the Gaussian smoothing kernel in this research as: $\mathcal{W}(R';R)=\frac{1}{(2\pi)^{3/2}R^3} \exp\left(-\frac{R'^2}{2R^2}\right)$. For the sake of convenience, we will omit the subscript "smoothed" from $\delta_{\rm smoothed}^{\diamond}(\boldsymbol{r}, t)$ and will assume the use of the smoothed field unless otherwise noted. Throughout this research, we take different snapshots for matter density field evolution in simulation labeled  by different redshifts as: $z=\{0,\, 0.5,\, 1.0,\, 2.0,\, 3.0\}$.  To perform the filtration process, we follow the steps introduced in \cite{2023arXiv231113520J}. Therefore  our cubical filtration method is super-level filtration, and the maximum threshold equates to the maximum value of $\nu^{\diamond}\equiv \log(\delta^{\diamond}+1)$ obtained in the mock data. Our cubic lattice is knows as cubical complex, and constructed $\delta^{\diamond}$ (Eq. (\ref{eq:delta})) can be interpreted as $\delta^{\diamond}: \mathcal{M}^{\diamond} \rightarrow \mathbb{R}$, which $\mathcal{M}^{\diamond} \subset \mathbb{R}^3$ and $\mathcal{M}^{\diamond}$ is a topological space related to ${\diamond}$ space. Utilizing  the super-level filtration, one can construct the excursion set as $\mathcal{M}^{\diamond}_{\nu}(z) = \{\boldsymbol{r} \in \mathcal{M}^{\diamond}(z) \mid \delta^{\diamond}(\boldsymbol{r},z) \geq \nu^{\diamond}\}$. Different excursion sets for redshift and real spaces are also generated by varying the corresponding thresholds as: $\nu^{\diamond}_{max} \geq \dots \geq \nu^{\diamond}_{i+1} \geq \nu^{\diamond}_{i} \geq \dots \geq \nu^{\diamond}_{min}$, or equivalently: $\mathcal{M}^{\diamond}_{\nu^{\diamond}_{max}}(z) \subseteq \dots \mathcal{M}^{\diamond}_{\nu^{\diamond}_{i+1}}(z) \subseteq
\mathcal{M}^{\diamond}_{\nu^{\diamond}_{i}}(z) \subseteq \dots \mathcal{M}^{\diamond}_{\nu^{\diamond}_{min}}(z) $ (for other types of filtration one can refer to \citep{2024arXiv240313985Y}. Also for more details about the $k$-Homology group of a generic topological space see Appendices A and B.  Figure ~\ref{fig:exset} illustrates the excursion sets generated at different threshold levels through the application of super-level filtration on a matter density field in real space, derived from a fiducial realization of the Quijote simulations. We have smoothed the simulated field with a Gaussian window function with smoothing scale $R=5$  Mpc $h^{-1}$. The upper and lower panels in mentioned figure are snapshots for $z=0$ and $z=2$, respectively.
\begin{figure*}
    \begin{center}
        \includegraphics[scale=0.5]{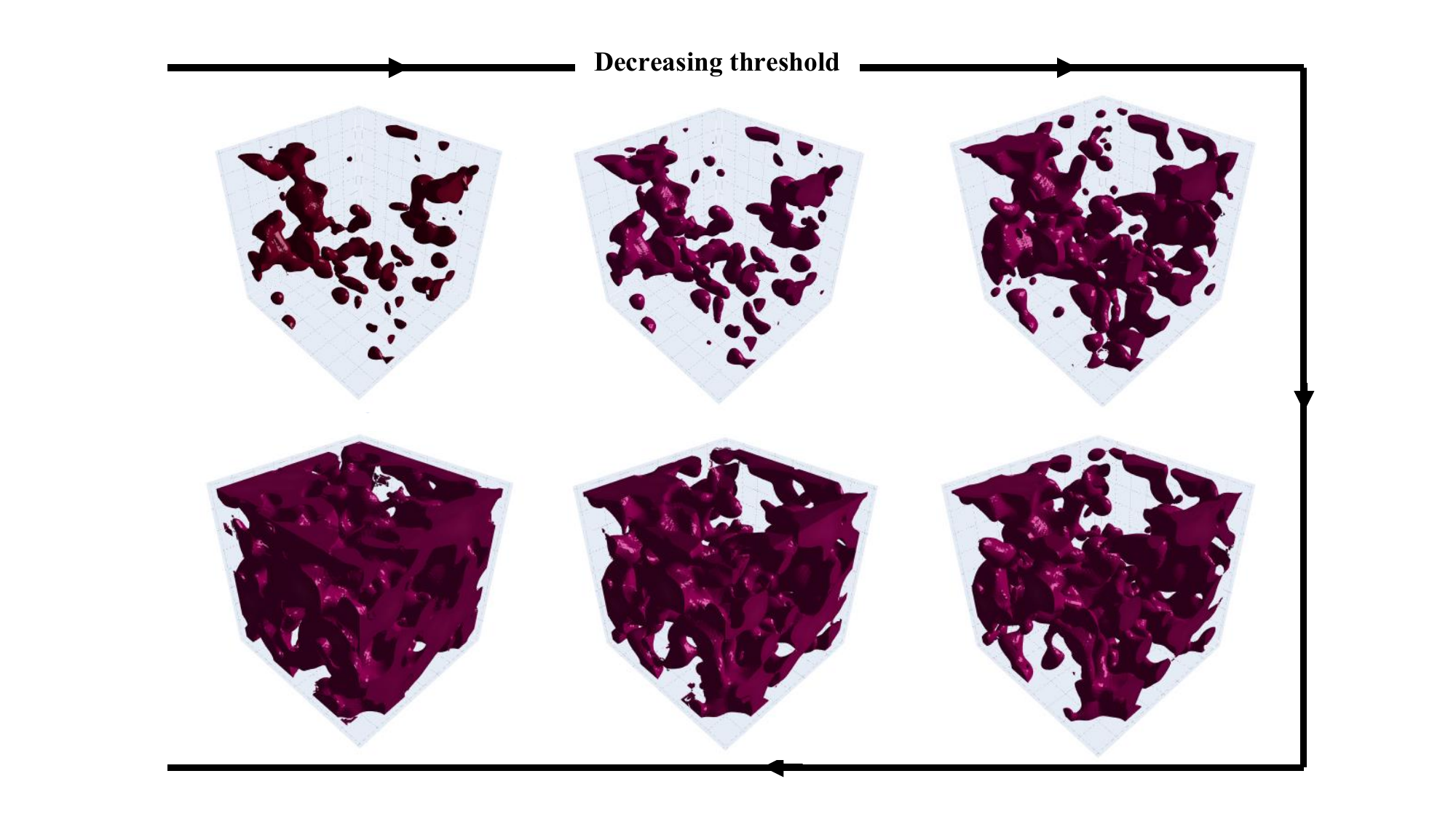}
        \includegraphics[scale=0.5]{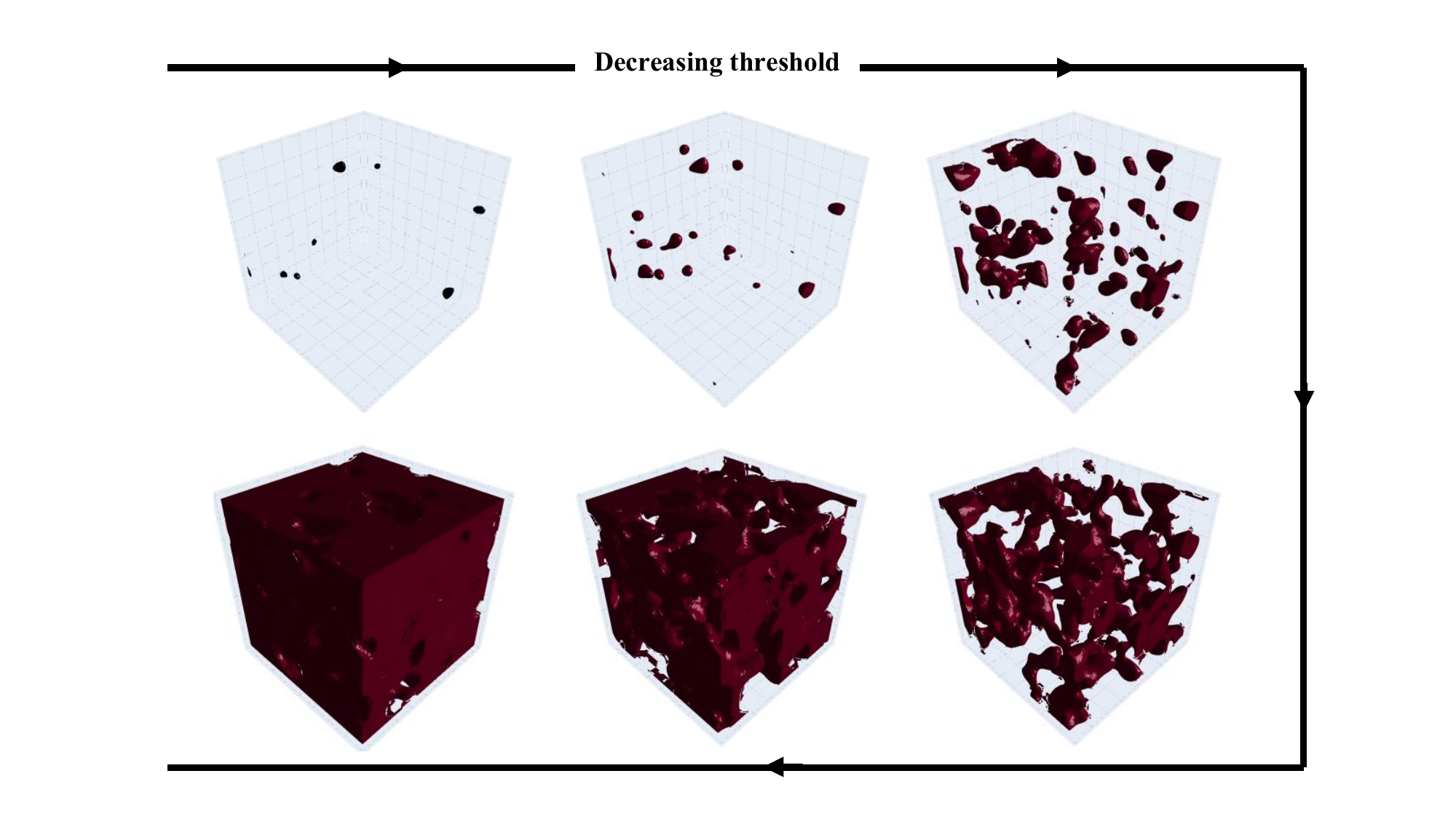}
    \end{center}
    \caption{Constructed excursion sets by implementation of super-level filtering on the density field in the real space using a sample from Quijote simulations.  For better illustration, we cut a box of $156$ Mpc $h^{-1}$ size from the original volume and utilized the cloud-in-cell scheme by Pylians. We take threshold levels as $\nu = [1.3, 1.0, 0.5, 0.3, 0.0, -0.2]$ and confidently smoothed the density field using a Gaussian window function with a smoothing scale of $R= 5$ Mpc $h^{-1}$. Upper  and lower panels are for  $z=0$ and $z=2$, respectively. The arrow depicts the direction of decreasing value of threshold.}
    \label{fig:exset}
\end{figure*}

\begin{table}[H]
    \centering
\begin{tabular}{|c|c|c|}
\hline
            Betti Number & Algebraic Topology & Cosmology \\
            \hline
            $\beta_0$ & Connected Components & Clusters \\
            \hline
            $\beta_1$ & 1-dimensional Holes & Filament loops \\
            \hline
            $\beta_2$ & 2-dimensional Holes & Cosmic voids \\
            \hline
        \end{tabular}
        \caption{\label{table11} Definition of Betti numbers in cosmology. }
    \end{table}

    The objective of Persistent Homology is to find structure in data and extract topological information of $\mathcal{M}$ via measuring lifetime of topological features and studying topological invariants \cite{nakahara2003geometry,munkres2018elements}. The topological properties of $\mathcal{M}^{\diamond}_{\nu^{\diamond}_{i}}(z)$ are characterized by their Homology group as $\mathcal{H}_k^{\diamond}(\mathcal{M}^{\diamond}_{\nu^{\diamond}}(z))$ and for 3-dimensional embedded cubical complex, $k=0,\;1,\;2$. The rank of Homology group is determined by Betti number as $\beta^{\diamond}_k(\mathcal{M}^{\diamond}_{\nu^{\diamond}}(z))=|\mathcal{H}_k^{\diamond}(\mathcal{M}^{\diamond}_{\nu^{\diamond}}(z))|$.  The $\beta^{\diamond}_0$, $\beta^{\diamond}_1$ and $\beta^{\diamond}_2$ are respectively associated with the number of components, the number of independent loops, and the number of independent closed surfaces  (for more details see \citep{2017MNRAS.465.4281P,pranav2019topology,2019A&A...627A.163P,edelsbrunner2022computational,2021MNRAS.507.2968W}). In Table 1, we list the analogous of Betti numbers in algebraic-topology and cosmology.  More specifically, the association between Betti numbers and their implications for cosmological interpretations is influenced by both the nature of the filtration process and the underlying field.  Recently, Xu et al. have introduced a comprehensive method referred to as the Significant Cosmic Holes in Universe, designed to effectively identify the distinct elements of the cosmic web  \citep{Xu2019}.

    Topological features evolve by varying the threshold from $\mathcal{M}^{\diamond}_{\nu^{\diamond}_{i+1}}(z)$ to $\mathcal{M}^{\diamond}_{\nu^{\diamond}_{i}}(z)$ according to performing super-level filtration process. The thresholds associated with the appearance and disappearance of a $k$-hole are noted by means of persistent pairs as $\nu^{\diamond-(k)}=\left(\nu^{\diamond-(k)}_{birth}, \nu^{\diamond-(k)}_{death}\right)$ and a multiset, $\mathcal{D}^{\diamond}_k=\{\nu^{\diamond-(k)}_i\}$, is known as persistence diagram. To construct tractable and interpretable topological features, we assume a summary statistics including following quantities computed by  $\mathcal{D}^{\diamond}_k$ as: \\
    (1) The $k$th Betti number:
    \begin{equation}
    \beta^{\diamond}_{k}(\nu) ={\sum_{i=1}^{m^{\diamond}_k} } {\Theta}\left(\nu^{\diamond-(k)}_{(i),birth} - \nu\right) {\Theta}\left(\nu^{\diamond-(k)}_{(i),death} - \nu\right)
    \end{equation}
    where $\Theta$ indicates Heaviside function, and $m^{\diamond}_k$ is the total number of persistence pairs related to the $k$th persistence diagram for either redshift or real spaces.  In this study, we normalize the $\beta^{\diamond}_k$ by the volume of the box, which we denote as $\tilde{\beta}^{\diamond}_k \equiv \frac{\beta^{\diamond}_k}{Volume}$. Hereafter, we will refer to it as the reduced Betti number.\\
    (2) Area under the Betti number curve:
    \begin{equation}
    A^{\diamond}_k = \int d\nu\; \tilde{\beta}^{\diamond}_k (\nu)
    \end{equation}
    (3) The Persistent entropy: Inspired by Shannon entropy for a typical state probability, one can define the persistence entropy (PE) of the $k$th persistence barcode. We construct the probability for lifetime of homology classes at a given threshold as:
    \begin{equation}
    PE^{\diamond}_k =-{\sum_{i=1}^{m^{\diamond}_k} } {\frac{\ell^{\diamond-(k)}_i}{{\sum_{j=1}^{m^{\diamond}_k}} \ell^{\diamond-(k)}_j}}{\log\left(\frac{\ell^{\diamond-(k)}_i}{{\sum_{j=1}^{m^{\diamond}_k}} \ell^{\diamond-(k)}_j}\right)}
    \end{equation}
    here $\ell^{\diamond-(k)}_i$ is the lifetime of each persistence pairs and it is defined by ${\ell^{\diamond-(k)}_i} \equiv\left(\nu^{\diamond-(k)}_{(i),birth}- \nu^{\diamond-(k)}_{(i),death}\right)$  \cite{masoomy2021persistent}.  Subsequently, our feature vector which is also known as summary statistics becomes ${\rm PH}^{\diamond}:\left(\tilde{\beta}^{\diamond}_{k},\;  A^{\diamond}_k,\;  PE^{\diamond}_k   \right)$. In addition to mentioned feature vector, it worth mentioning that  various methods have been introduced to map the persistence diagram to summary statistics \citep{adams2017persistence}. Since our data (matter density field) is 3-dimension and it looks like an image, we use Cubical Ripser \cite{kaji2020cubical} for computing persistence diagrams. In the following parts, we will compute the ${\rm PH}^{\diamond}$ for both real space and distorted field by RSD phenomenon to examine the impact of RSD on the matter density field for different smoothing scales and various redshifts. Accordingly, we quantify the sensitivity of our summary statistics to the linear Kaiser and non-linear RSD.

    \begin{figure}
        \begin{center}
            \includegraphics[scale=0.3]{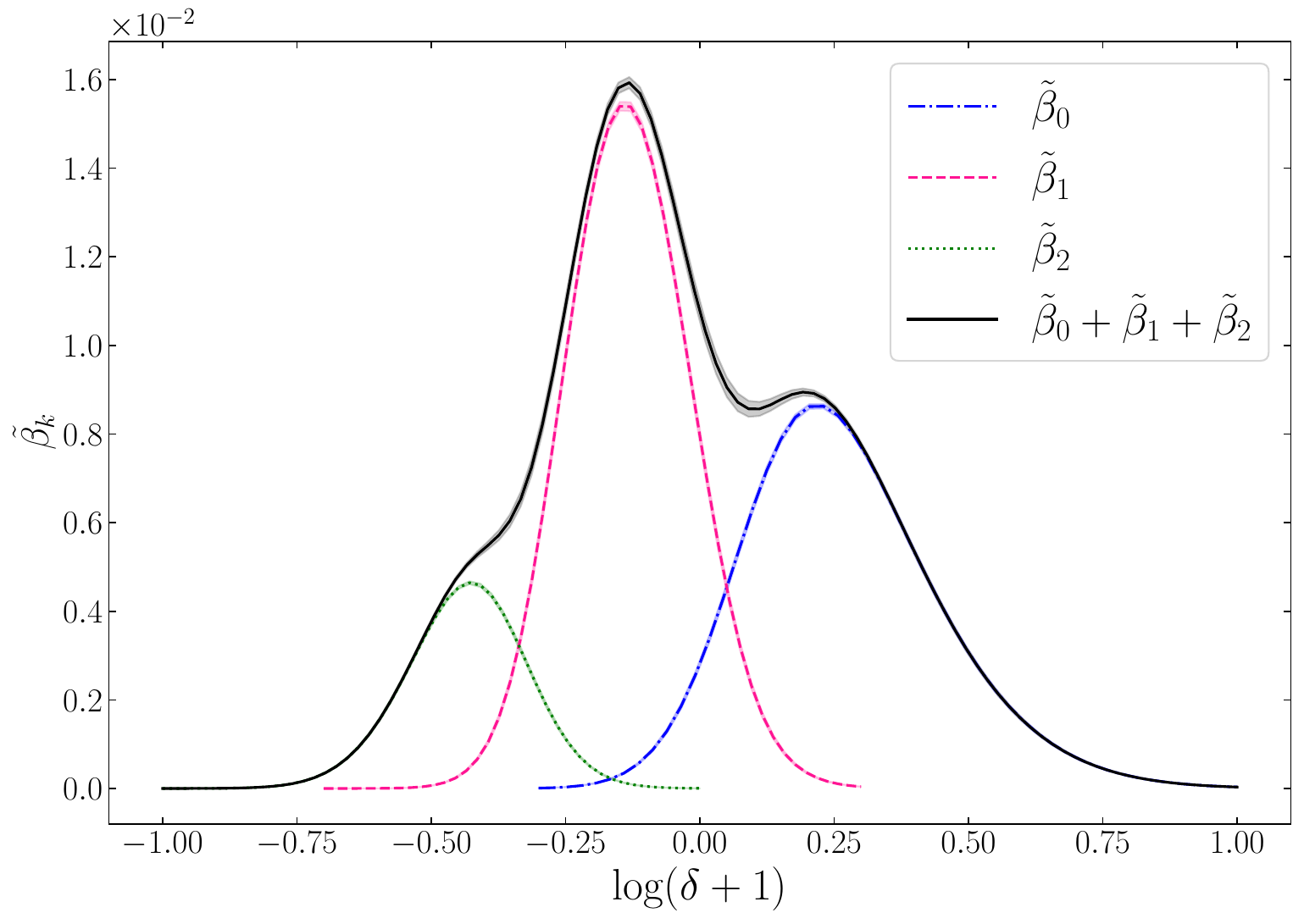}
            \caption{The reduced  Betti number curves for different $k$-holes as a function of threshold for smoothed matter density field simulated by Quijote. The black solid line reveals the summation of $\tilde{\beta}$s. The smoothing scale is $R = 5$ Mpc $h^{-1}$ and the redshift is $z = 0$. The error-bar at $2\sigma$ level of confidence is represented by shaded area around each curve.}
            \label{fig:bcsR5real}
        \end{center}
    \end{figure}

    \section{Description of synthetic data sets} \label{sec:data}
    In order to investigate the matter density field within redshift space and assess the effects of distortions on the corresponding topology, we employ the Quijote N-body simulations suite. This suite utilizes the Three+SPH code GADGET-III and comprises a total of 44,100 N-body realizations. Each realization is composed of $512^3$ dark matter particles, arranged within a cubic volume of ${1\,\text{Gpc}^3\,h^{-3}}$ with periodic boundary conditions, based on the designated fiducial parameters:
    $\{w=-1,\,{\sigma_8}=0.834, n_s=0.9624, h=0.6711,\,  {\Omega_b}=0.049,\, {\Omega_m}=0.3175\}$ for real space  \cite{aghanim2020planckvi}. To generate the corresponding redshift space, we use plane-parallel approximation according to Eqs. (\ref{eq:redshift position}) and (\ref{eq:anisotropic delta}), taking $b=1$, we obtain  $\mathcal{B}=0.526$ at current epoch for mentioned values of cosmological parameters.

    We then make density contrast field by using Pylians \cite{villaescusa2018pylians} routine and considering  CIC \footnotetext{Cloud-in-Cell (a mass-assignment scheme)} method.  We are also interested in examining the redshift and smoothing scale dependency of ${\rm PH}^{\diamond}$. Therefore, we create  smoothed matter density fields for snapshots data in different redshifts $z= \{0.0,\; 0.5,\; 1.0,\; 2.0,\; 3.0\}$ and taking the Gaussian smoothing function with different smoothing scales $5\,\;\text{Mpc}\,\;h^{-1} \le R\,\le 45\,\;\text{Mpc}\,\;h^{-1}$.

    The minimum value of $R$ is selected to be $R = 5\,\;\text{Mpc}\,\;h^{-1}$ which is larger than the voxel size of this simulation and the impact of shot-noise becomes  diminish \cite{2023arXiv231113520J}. In this research, we make 100 realizations of matter density field for each redshift and smoothing scale by using the Quijote large-scale structure simulation data for our numerical calculations.

    \begin{figure}
        \begin{center}
            \includegraphics[scale=0.3]{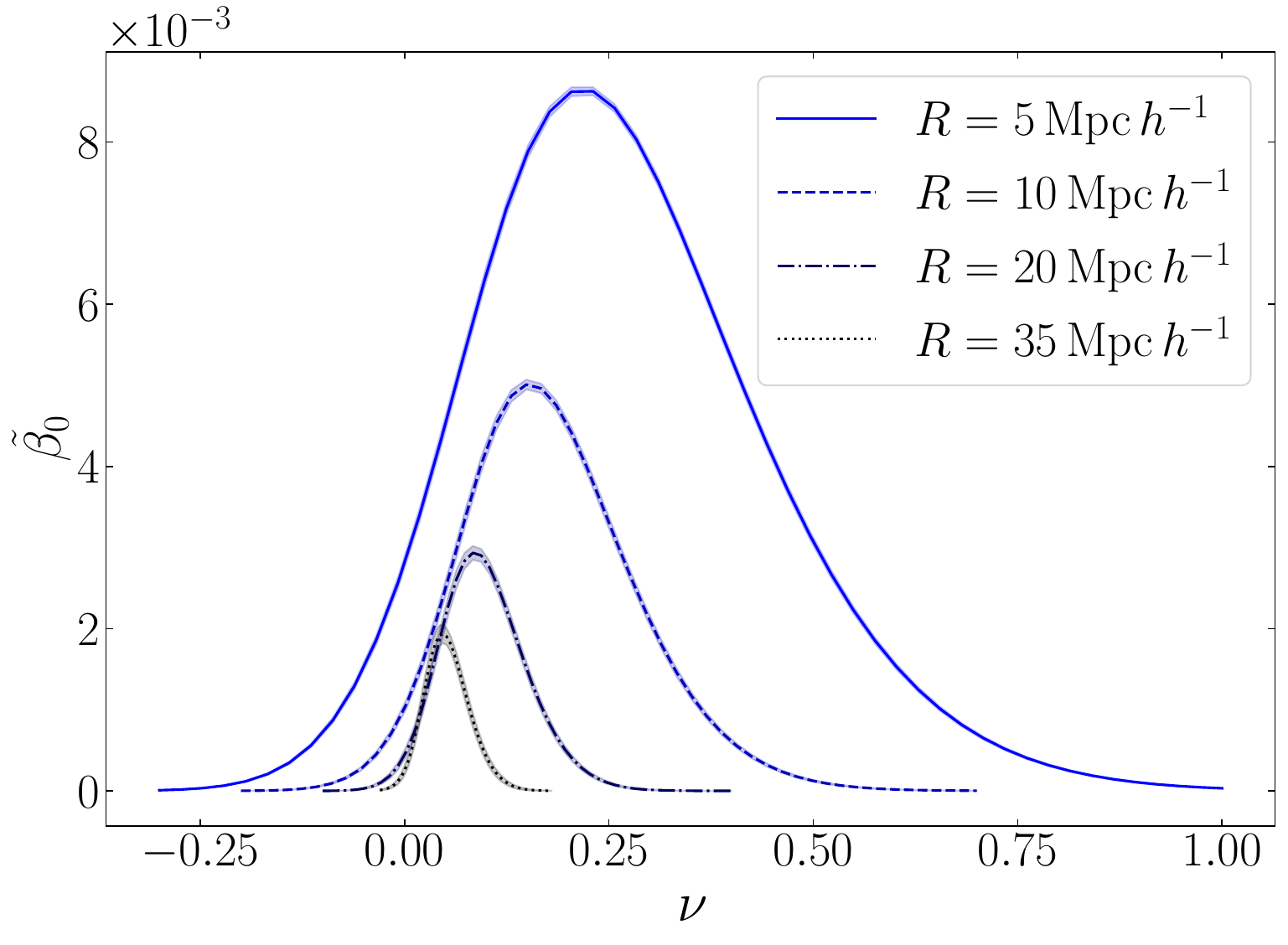}
            \includegraphics[scale=0.3]{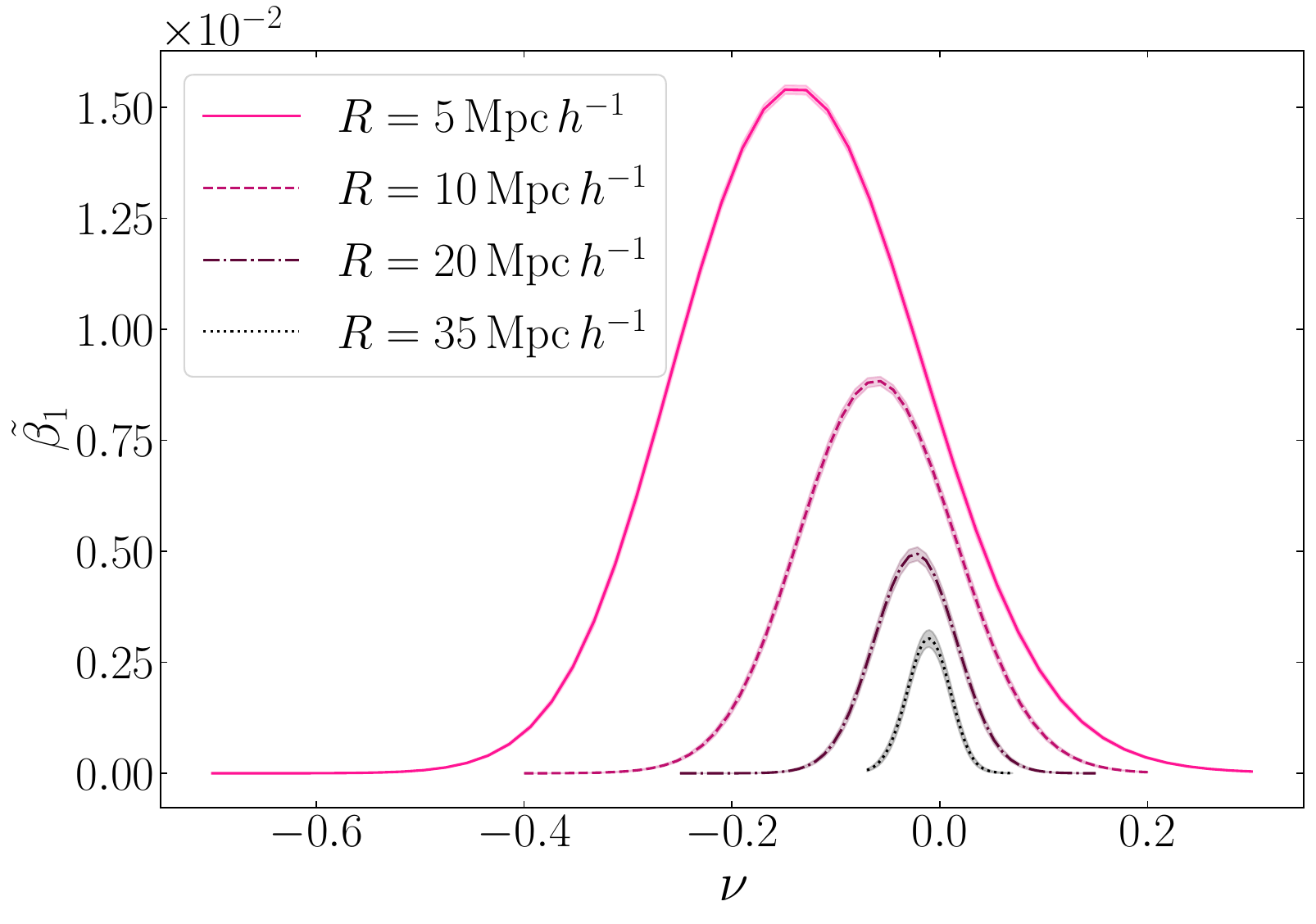}
            \includegraphics[scale=0.3]{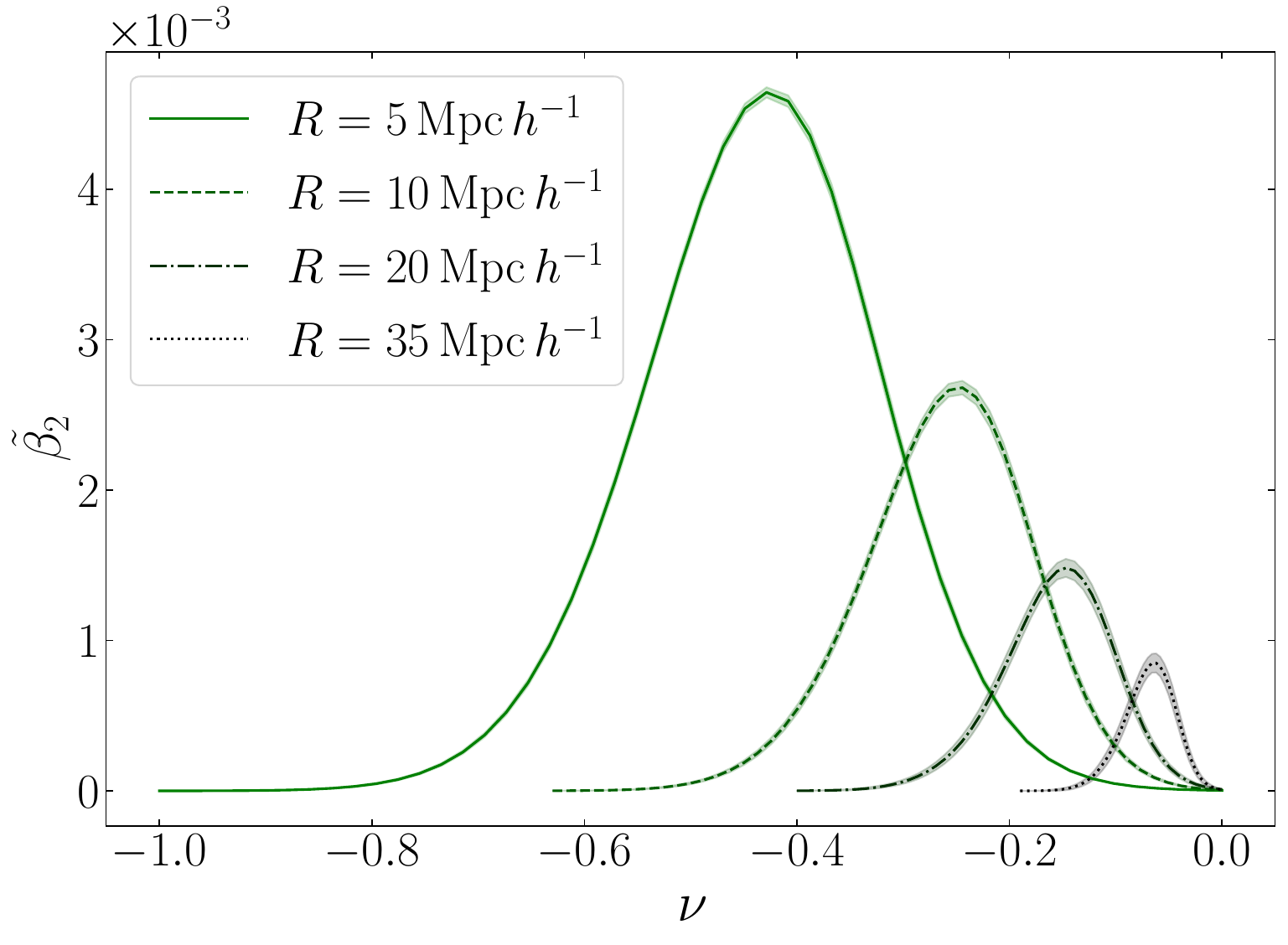}
        \end{center}
        \caption{Evolution of the reduced Betti number curves with respect to different smoothing scales for synthetic matter density field in the real space at $z=0$. Upper, middle and lower panels correspond to $\tilde{\beta}_0$, $\tilde{\beta}_1$ and $\tilde{\beta}_2$, respectively. The error-bar has been computed at $2\sigma$ level of confidence and depicted by shaded area around each curve. }
        \label{fig:bcsreal}
    \end{figure}
    \begin{figure}
        \begin{center}
            \includegraphics[scale=0.3]{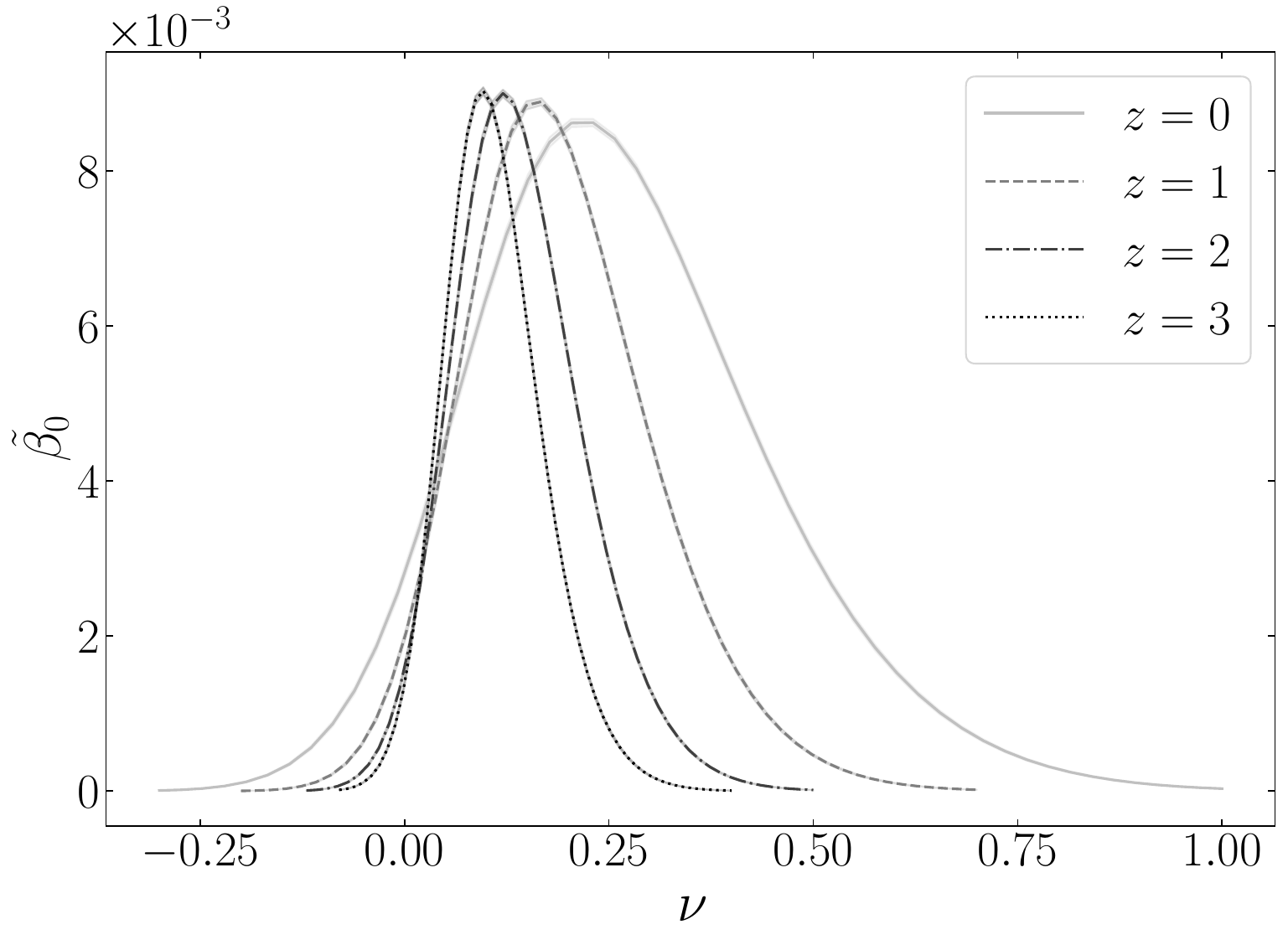}
            \includegraphics[scale=0.3]{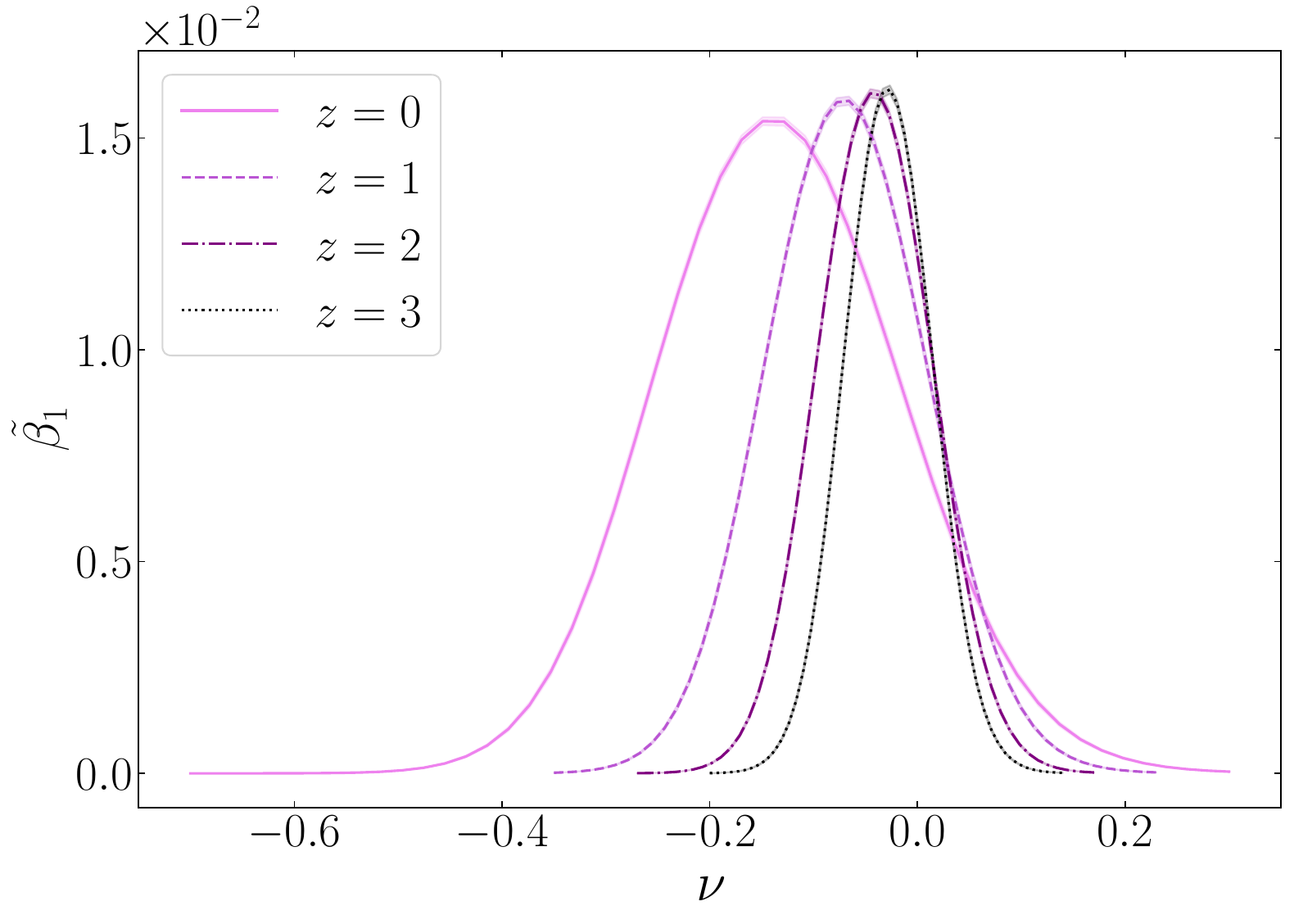}
            \includegraphics[scale=0.3]{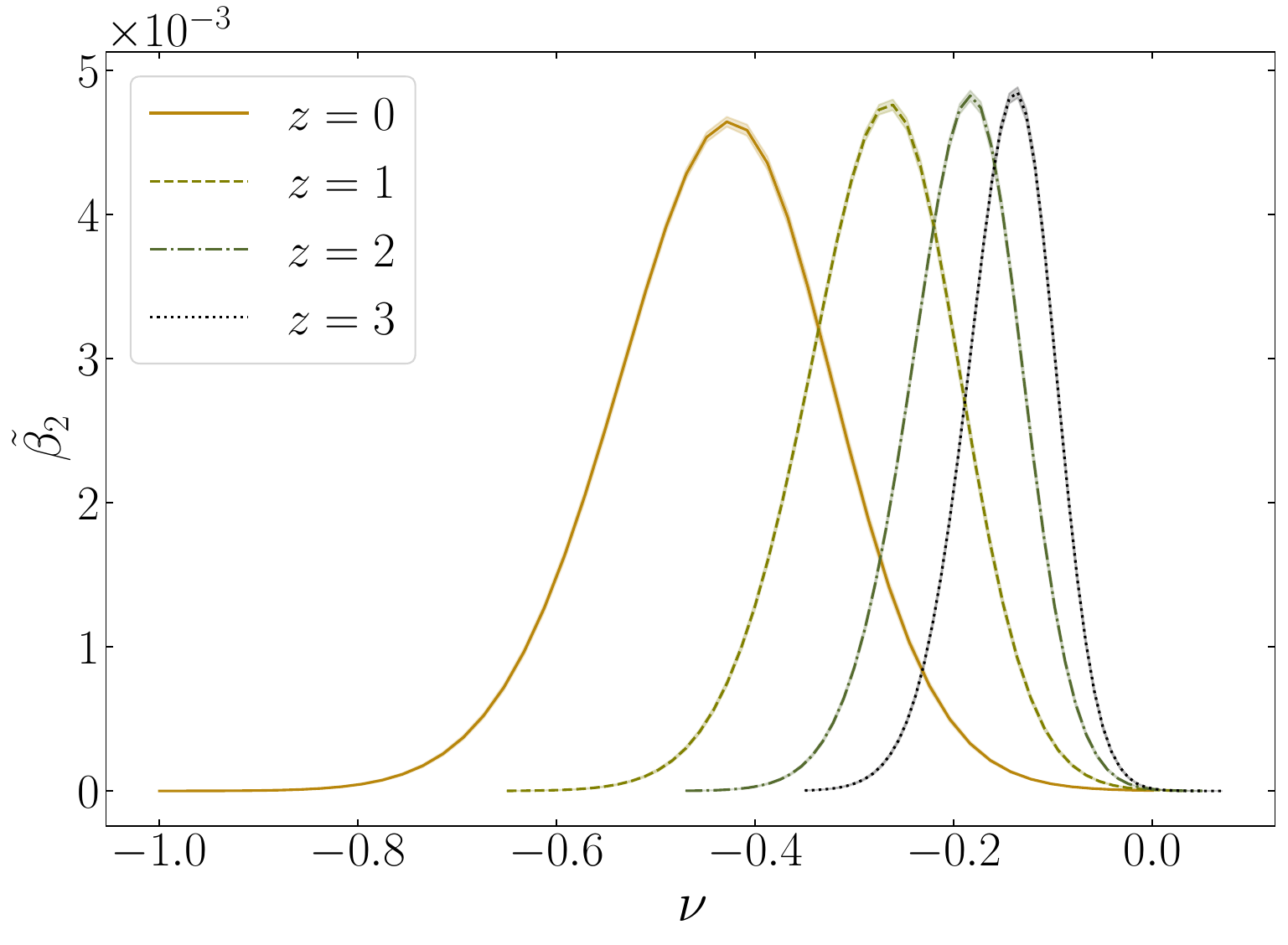}
        \end{center}
        \caption{The reduced Betti number curves of smoothed matter density field in real space with $R=5$ Mpc $h^{-1}$ for four different redshifts. Upper, middle and lower panels correspond to $\tilde{\beta}_0$, $\tilde{\beta}_1$ and $\tilde{\beta}_2$, respectively. The error-bar has been computed at $2\sigma$ level of confidence illustrated by shaded area around each curve. }
        \label{fig:bcsrealz}
    \end{figure}
    \section{Implementation of PH vectorization on the N-body Simulation }\label{sec:imp}
    In this section, we study the effect of redshift space distortions on the PH of the matter density field using the numerical derivation of our introduced summary statistics,
    ${\rm PH}^{(s,r)}:\left(\tilde{\beta}^{(s,r)}_{k},\;  A^{(s,r)}_k,\;  PE^{(s,r)}_k   \right)$.

    \subsection{PH of fiducial $\Lambda$CDM matter density field in real space }
    To check how the PH$^{(r)}$ vectorization behaves according to our pipeline (Fig. \ref{fig:pipeline}), we compute different elements of summary statistics for simulated matter density field by Quijote routine. Fig. \ref{fig:bcsR5real} depicts the $\tilde{\beta}_0$, $\tilde{\beta}_1$ and $\tilde{\beta}_2$ averaged over 100 realizations for fiducial $\Lambda$CDM matter density field in the real space. The underlying field has been smoothed by a Gaussian smoothing function with  $R= 5$ Mpc $h^{-1}$ at $z =0$. This figure demonstrates that the abundance of connected clusters ($k=0$) is almost alive at the positive threshold, while the $\tilde{\beta}_2$ that is analogously related to the void components persists in negative thresholds. The $2\sigma$ error associated with each case has been estimated by averaging over the independent realizations and it can be represented by the shaded region around each curve. The black solid line represents $\tilde{\beta}_0+\tilde{\beta}_1+\tilde{\beta}_2$.
    It confirms that Betti number give more information than their linear combination such as Euler characteristic.

    The impact of smoothing scale on the Betti number has been depicted in Fig.  \ref{fig:bcsreal}. The upper panel is devoted to evolution of $\tilde{\beta}_0$ for different values of smoothing scales, $R=5$ Mpc $h^{-1}$, $R=10$ Mpc $h^{-1}$, $R=20$ Mpc $h^{-1}$, $R=35$ Mpc $h^{-1}$. Increasing the value of $R$ leads to a deceasing in the value of $\tilde{\beta}_0$, as well as a transition of connected components from high threshold levels to the mean threshold value. The middle panel depicts the behavior of $\tilde{\beta}_1$ versus $\nu$ for different $R$. While the lower panel shows how the reduced abundance of closed surfaces (voids) is modified by increasing the value of smoothing scale. Comparing the evolution behavior of  $\tilde{\beta}_2$ with reduced Betti number for $k=0$ and $k=1$ demonstrates that the  imprint of smoothing scale on the maximum value of $\tilde{\beta}_2$ recognized by the super-level filtration method is higher than other reduced Betti number.  In conclusion, an increase in the parameter $R$ leads to a reduction in both the threshold interval and the amplitude of the reduced Betti number. It is important to emphasize that as the smoothing scales increase, the non-linear characteristics of the matter density field become less pronounced.

    In addition, the influence of redshift on the topological properties is also other key point that should be examined for matter density field in the real and redshift spaces. To realize the impact of redshift, we also extract the excursion sets for different cosmic redshifts, and the reduced Betti number are computed. Fig. \ref{fig:bcsrealz} indicates the $\tilde{\beta}_k$ for $0-$, $1-$ and $2-$holes as a function of threshold for different redshifts.  By increasing the redshift, the width of the abundance of topological component curves decreases and they are shifted along the $\nu$ axis toward the mean value of the threshold. This behavior can be explained by means of decreasing the rate of gravitational collapse by increasing redshift. This observation is also compatible with behavior of the combined mean of both holes and connected clusters defined by the eigenvalues of rank-two MT ($W^{1,1}_2$) reported in \cite{appleby2018minkowski}.  Unlike the contributions associated with the smoothing scale, our analysis reveals that the peak value of $\tilde{\beta}_k$ exhibits considerable robustness against raising redshift, at least for the interval analyzed in this investigation. Additionally, the dependence of $\tilde{\beta}_2$ on redshift demonstrates a greater sensitivity compared to other topological features.

    \subsection{PH of mock matter density field in redshift space}

    In the previous subsection, we computed the Betti number curves for matter density field in the real space. The imprint of smoothing scale and redshift have been exploited. Now, we will focus on determining the summary statistics for the mock field modified by redshift distortions for different smoothing scales and redshift interval. The motivation for taking into account different smoothing scales as mentioned before is looking for the sensitivity of our feature vector to non-linear part of RSD. Also the effect of redshift space distortions can be interpreted as linear and non-linear influences of Hubble flow distortion by peculiar velocity. Therefore to compare aforementioned phenomena, taking into account different $R$ is reasonable.

    \begin{figure*}
        \begin{center}
            \includegraphics[scale=0.23]{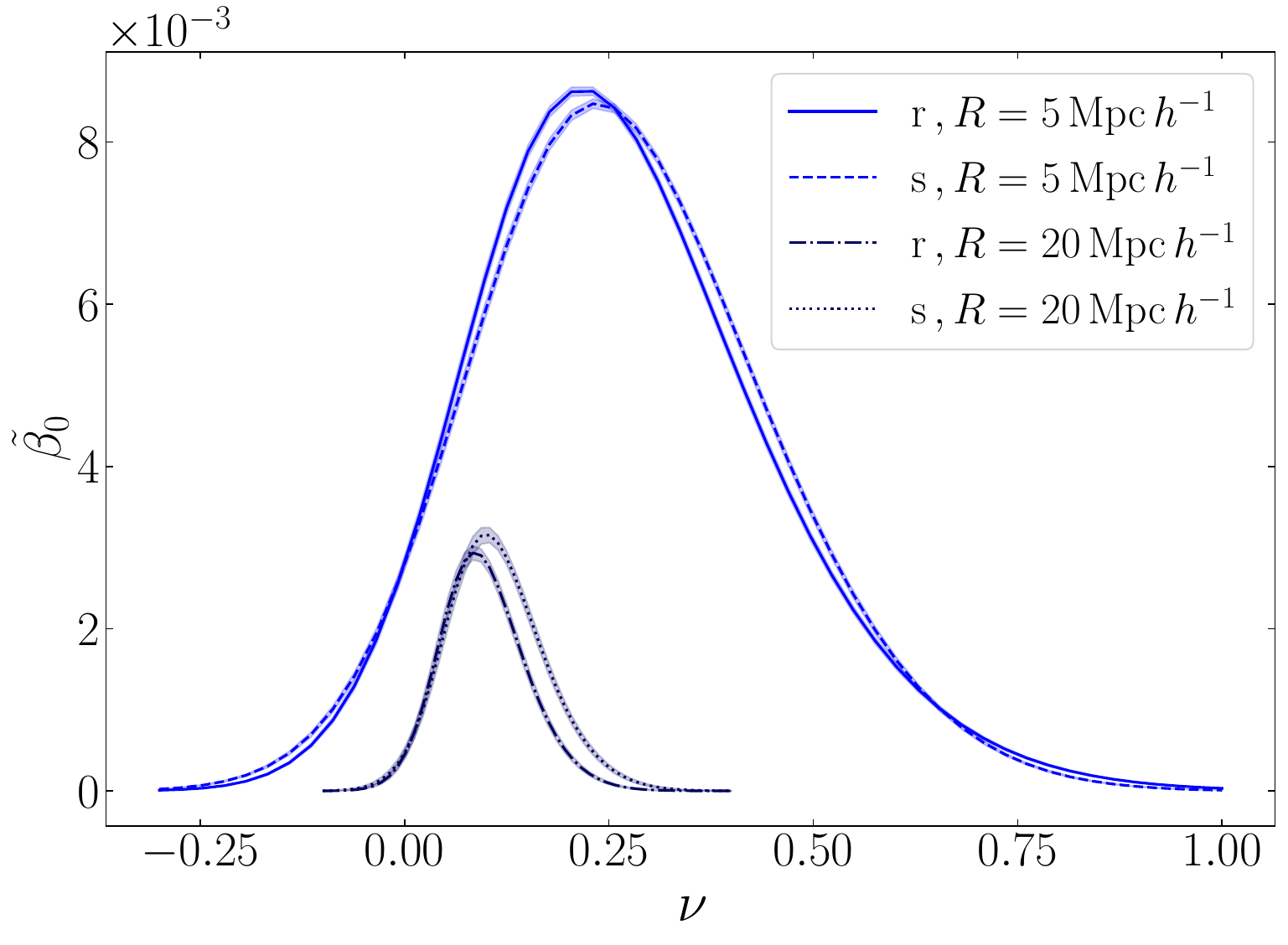}
            \includegraphics[scale=0.23]{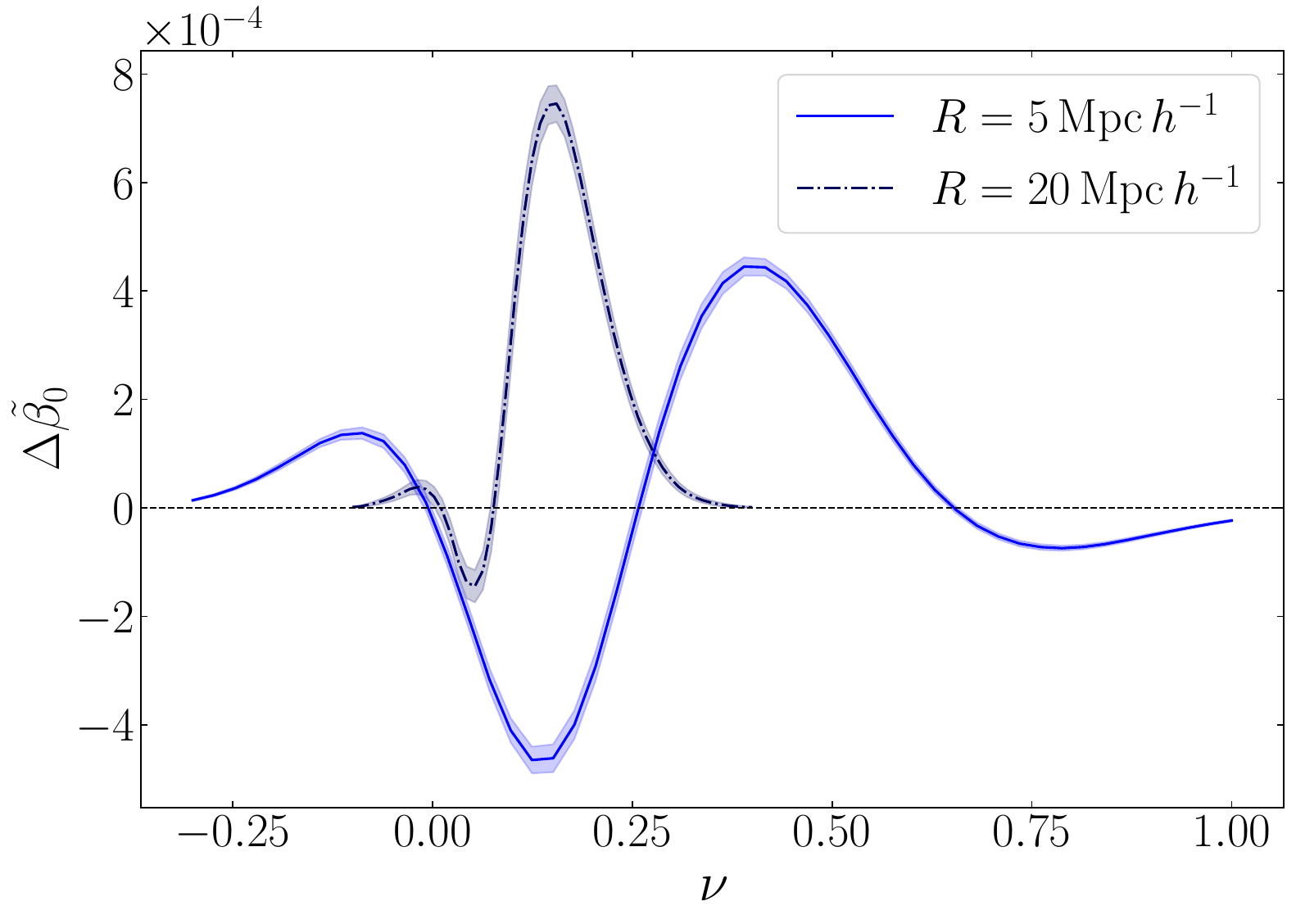}
            \includegraphics[scale=0.23]{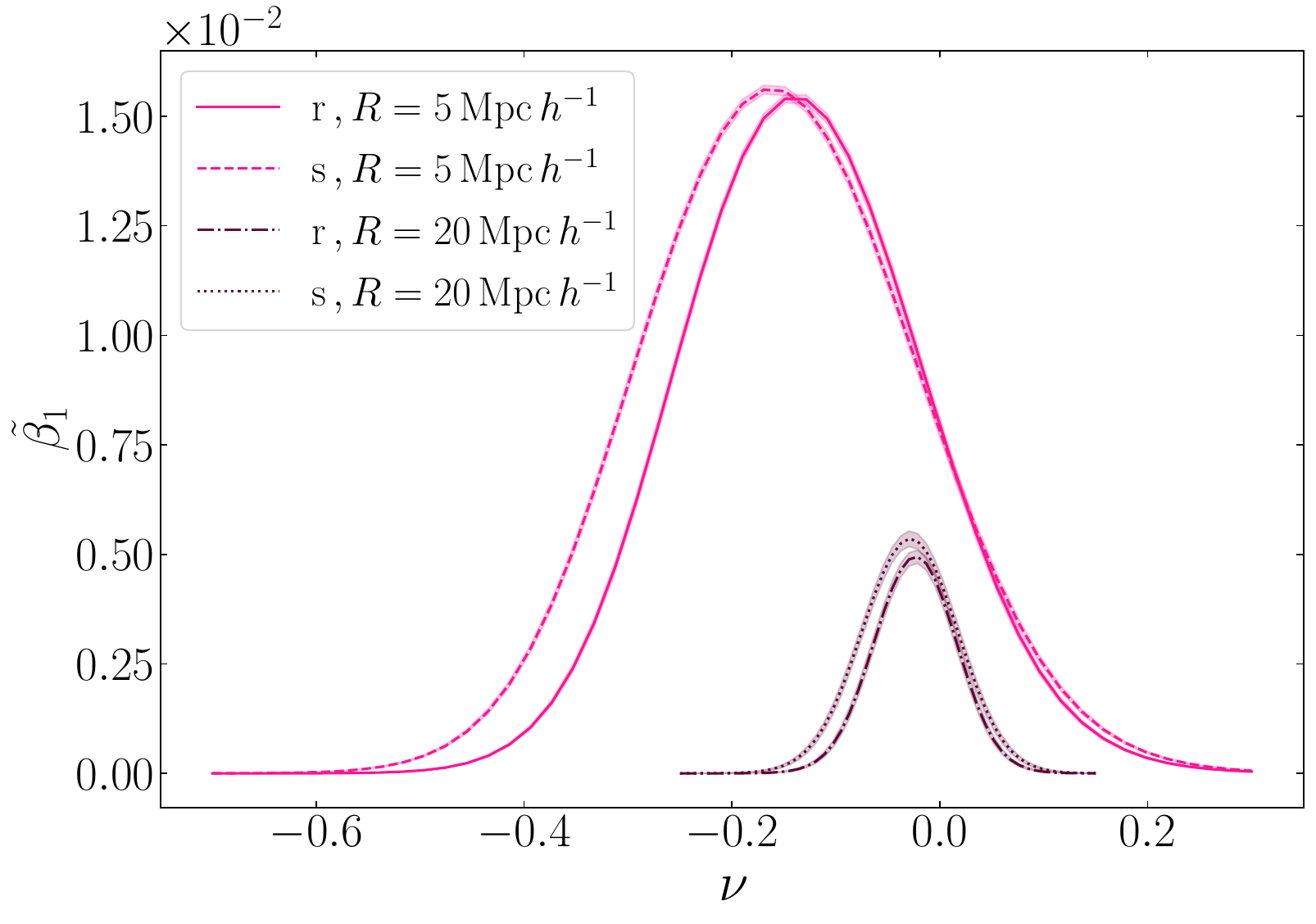}
            \includegraphics[scale=0.23]{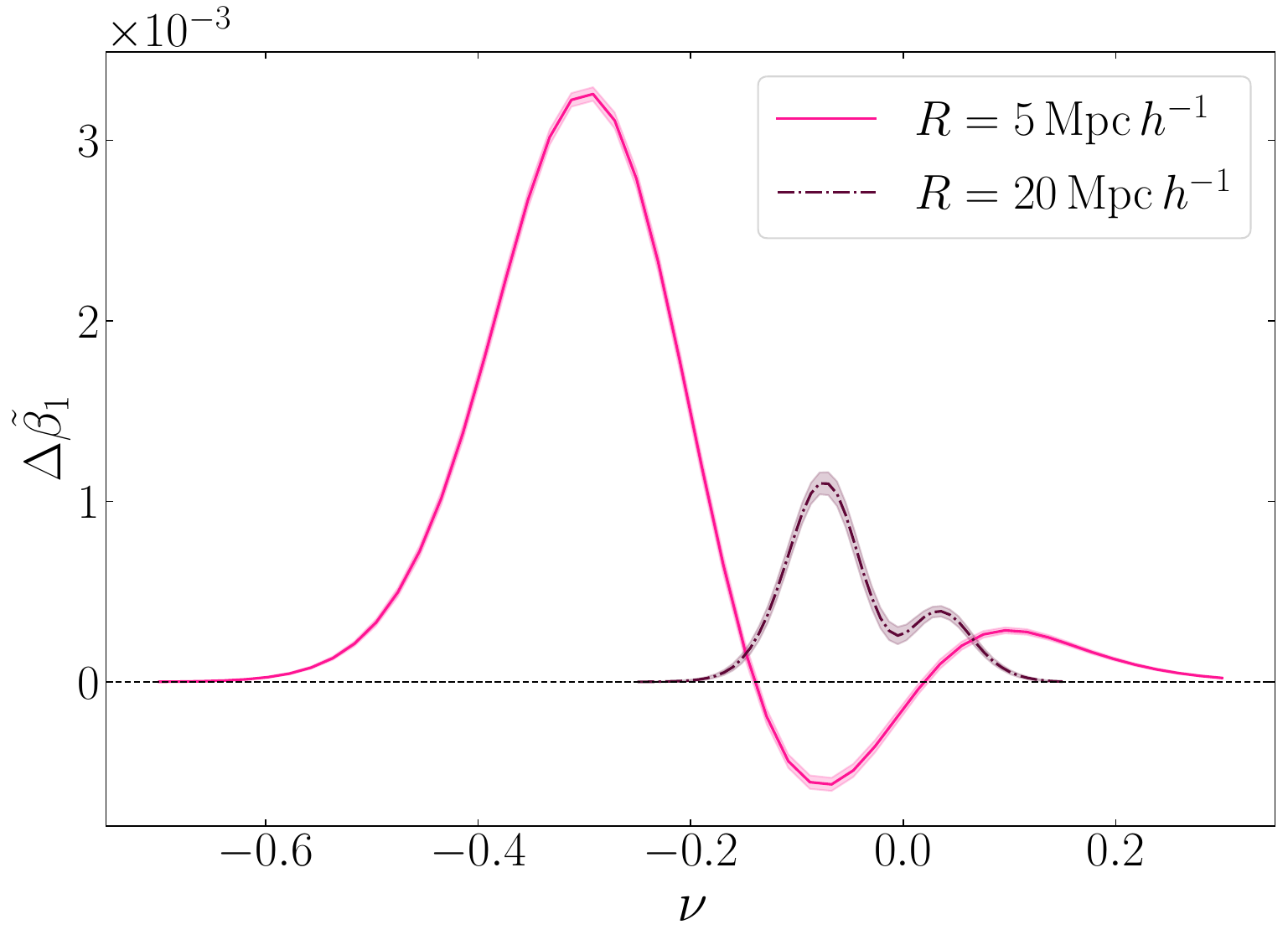}
            \includegraphics[scale=0.23]{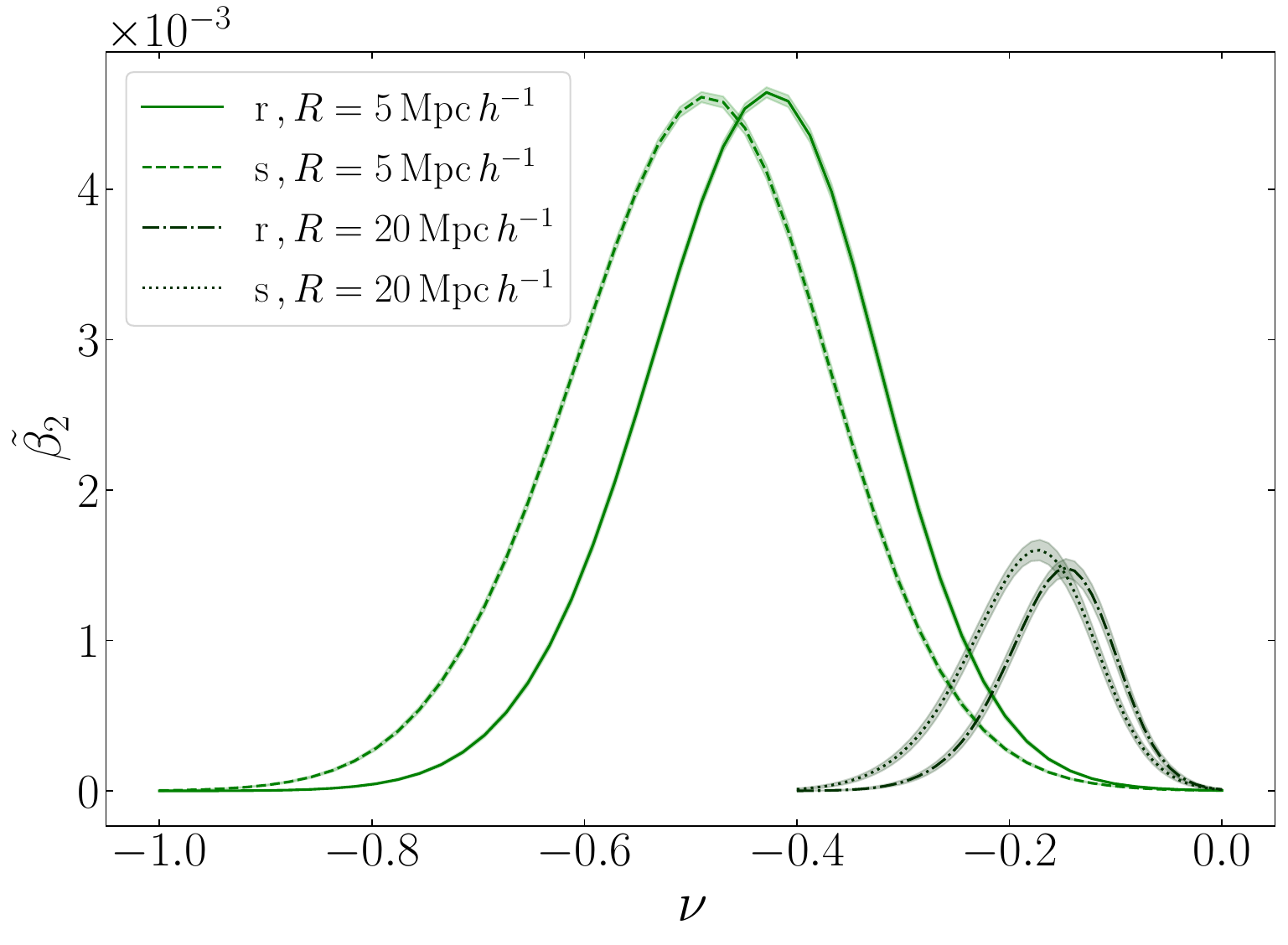}
            \includegraphics[scale=0.23]{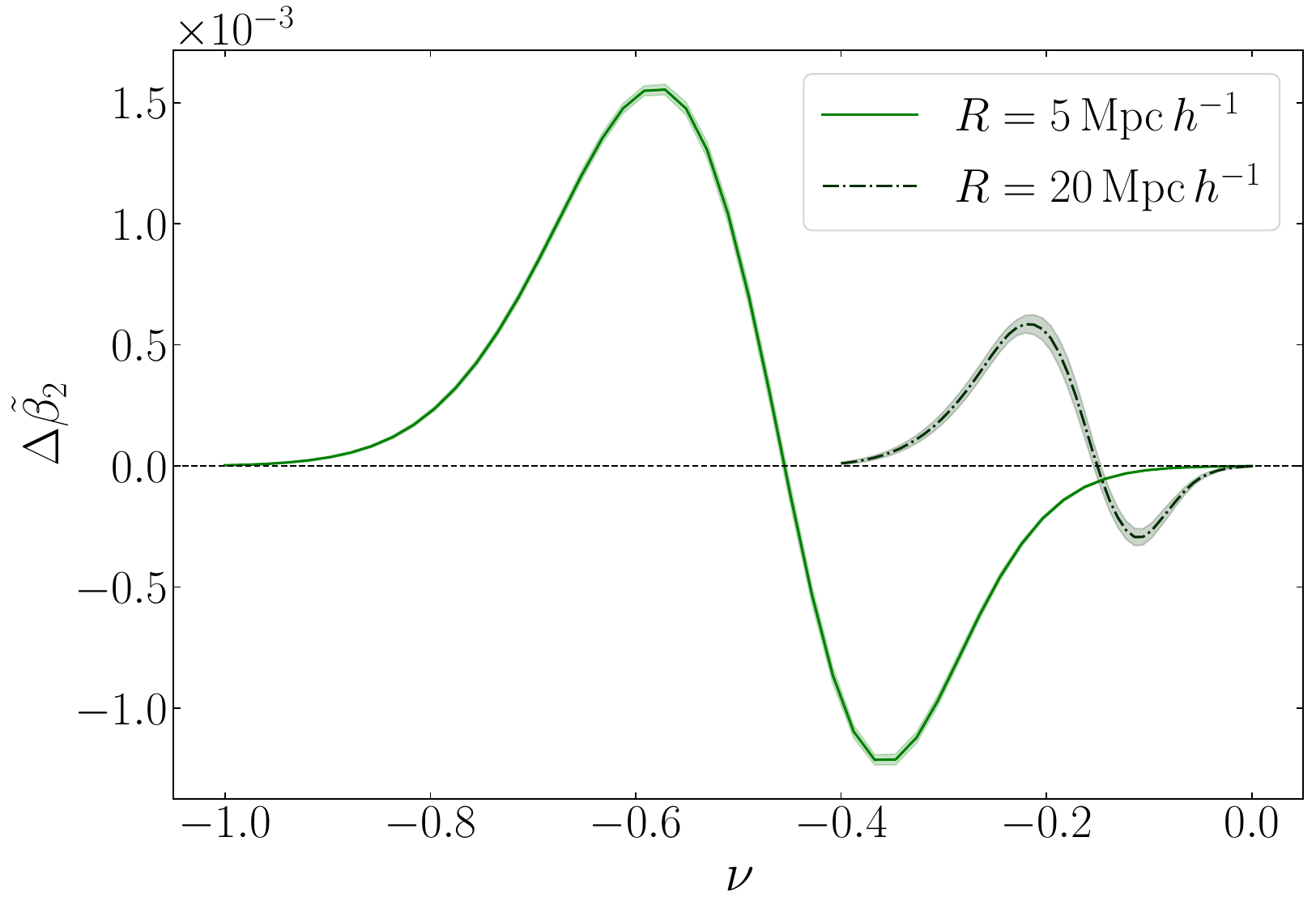}
        \end{center}
        \caption{The Betti number curves of smoothed matter density field  simulated in the real and redshift spaces for $z=0$ (left column). The $\Delta\tilde{\beta}_k\equiv \tilde{\beta}^{(s)}_k-\tilde{\beta}^{(r)}_k$ is given in right column. The error-bar has been computed at $2\sigma$ level of confidence and it is depicted by shaded area.}
        \label{fig:realredshift}
    \end{figure*}
    In Fig. \ref{fig:realredshift}, we  compare the Betti number for the synthetic matter density field at $z=0$ in both real space and associated redshift space for different smoothing scales. The overall response of RSD reveals an induced migration towards higher (lower) thresholds for clusters (filaments and voids).  Such behavior is justified as follows: for redshift space, the shape of over-dense regions are elongated (in small scales) or squashed (dominantly in large-scales) along the observer's line-of-sight. In addition according to Eq. (\ref{eq:anisotropic delta}), the density contrast has been modified by factor $\tilde{O}_s$, consequently, in our super-level filtration process,  we expect that the connected components population moves to the higher thresholds.  On the other hand, the population of filaments and voids  are shifted to the more negative thresholds when the matter density field is distorted by RSD (middle and lower rows of Fig. \ref{fig:realredshift}). Our results are consistent with other analysis for morphology of large-scale structures e.g. \cite{2022PhRvD.105j3028J}. The redshift space distortions are inevitable across all scales and therefore $\Delta \tilde{\beta}_k$ is not supposed to be zero unless for those scales and thresholds for which the linear Kaiser effect and the FoG effect cancel each other out. The occurrence of this compensation phenomenon might establish distinctive scales and thresholds that are useful for cosmological inferences, as highlighted in prior research \cite{2024ApJ...963...31K}.

    Comparing the  $\tilde{\beta}_0$ with independent loops and independent closed surfaces, we find that the clusters is less sensitive to RSD. While the $\tilde{\beta}_2$ is modified more than 0- and 1-holes. The $k$th holes encapsulate information pertaining to both linear and non-linear Redshift Space Distortion (RSD).

    \begin{figure*}
        \begin{center}
            \includegraphics[scale=0.24]{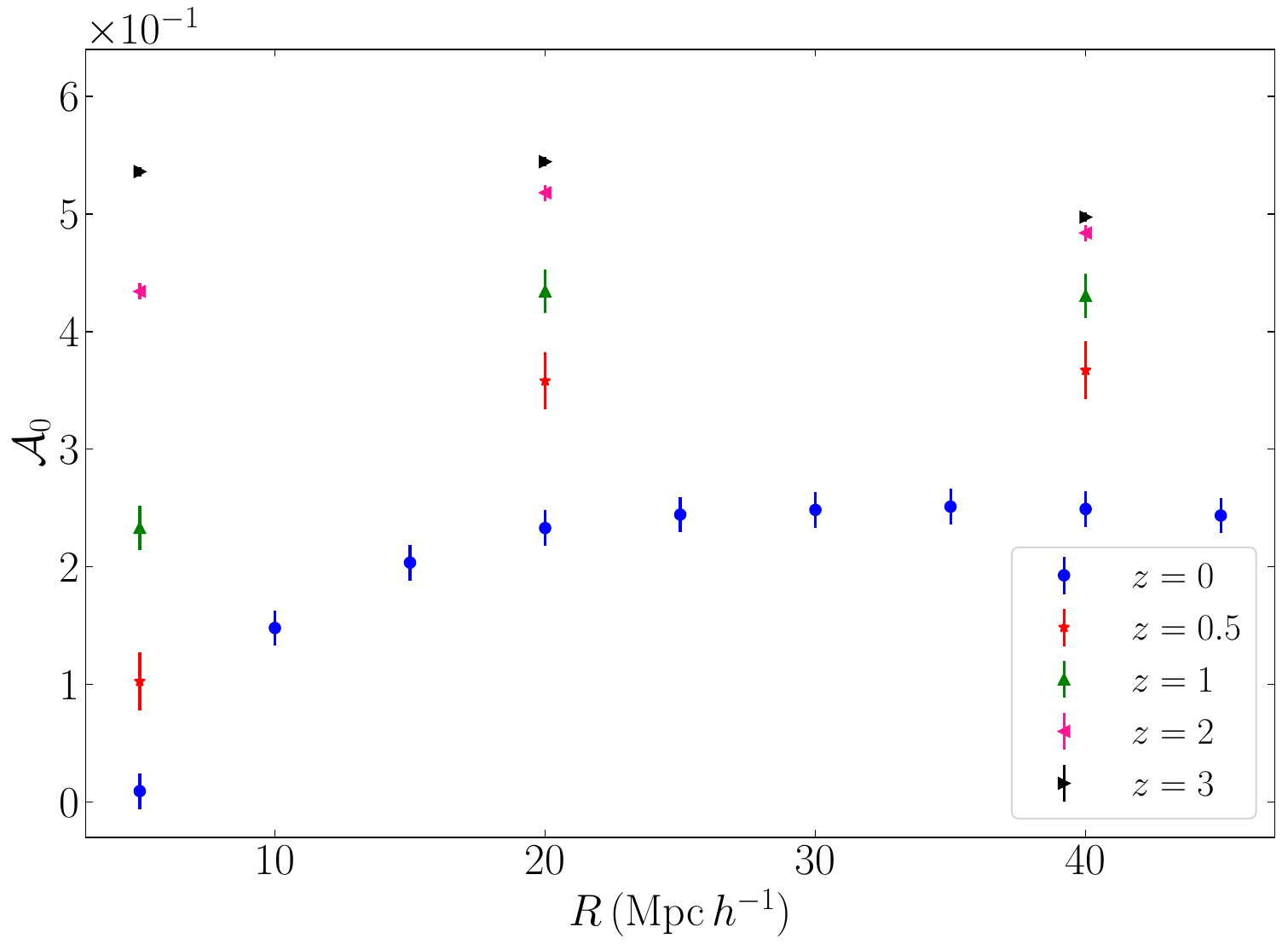}
            \includegraphics[scale=0.24]{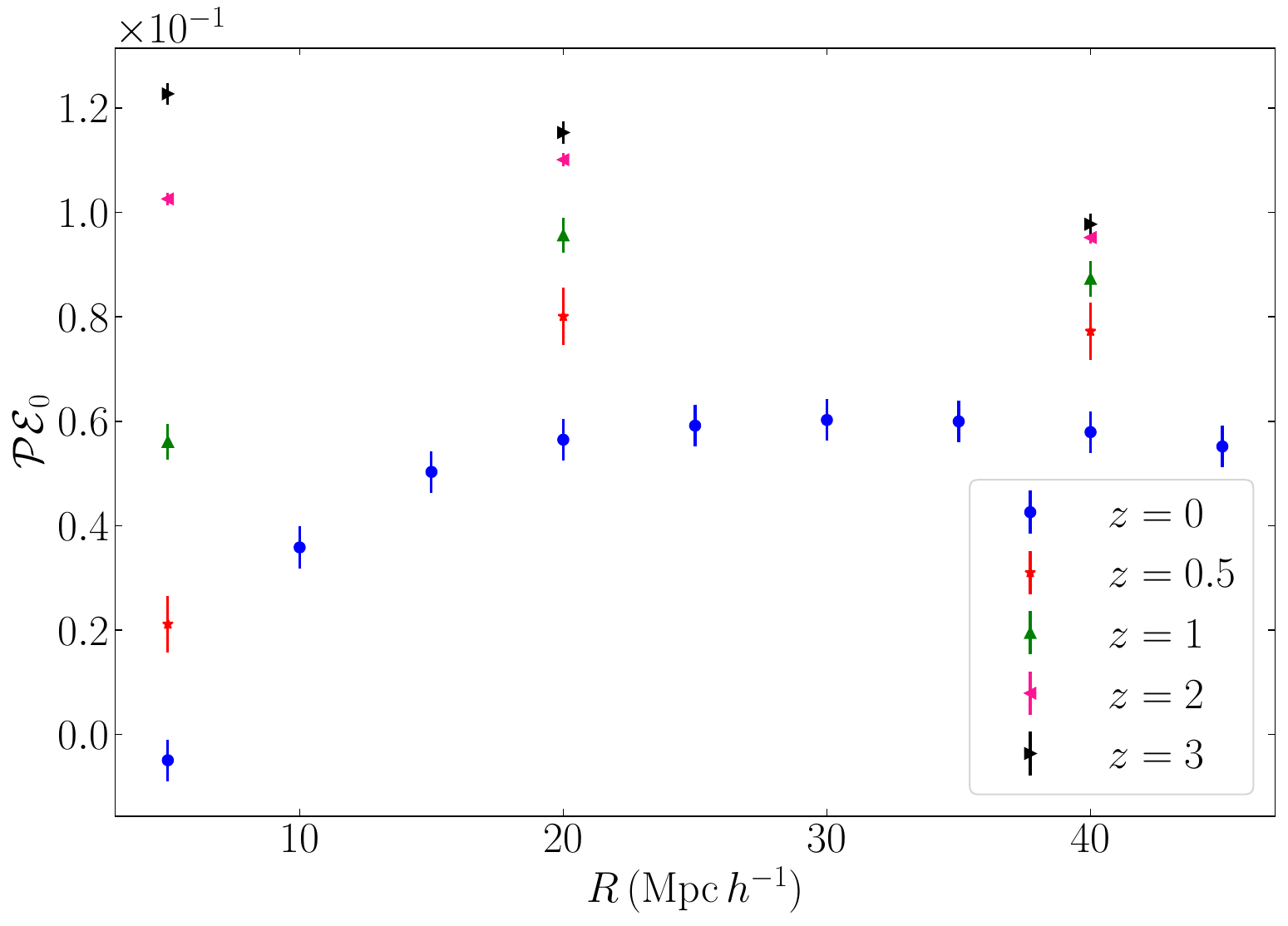}
            \includegraphics[scale=0.24]{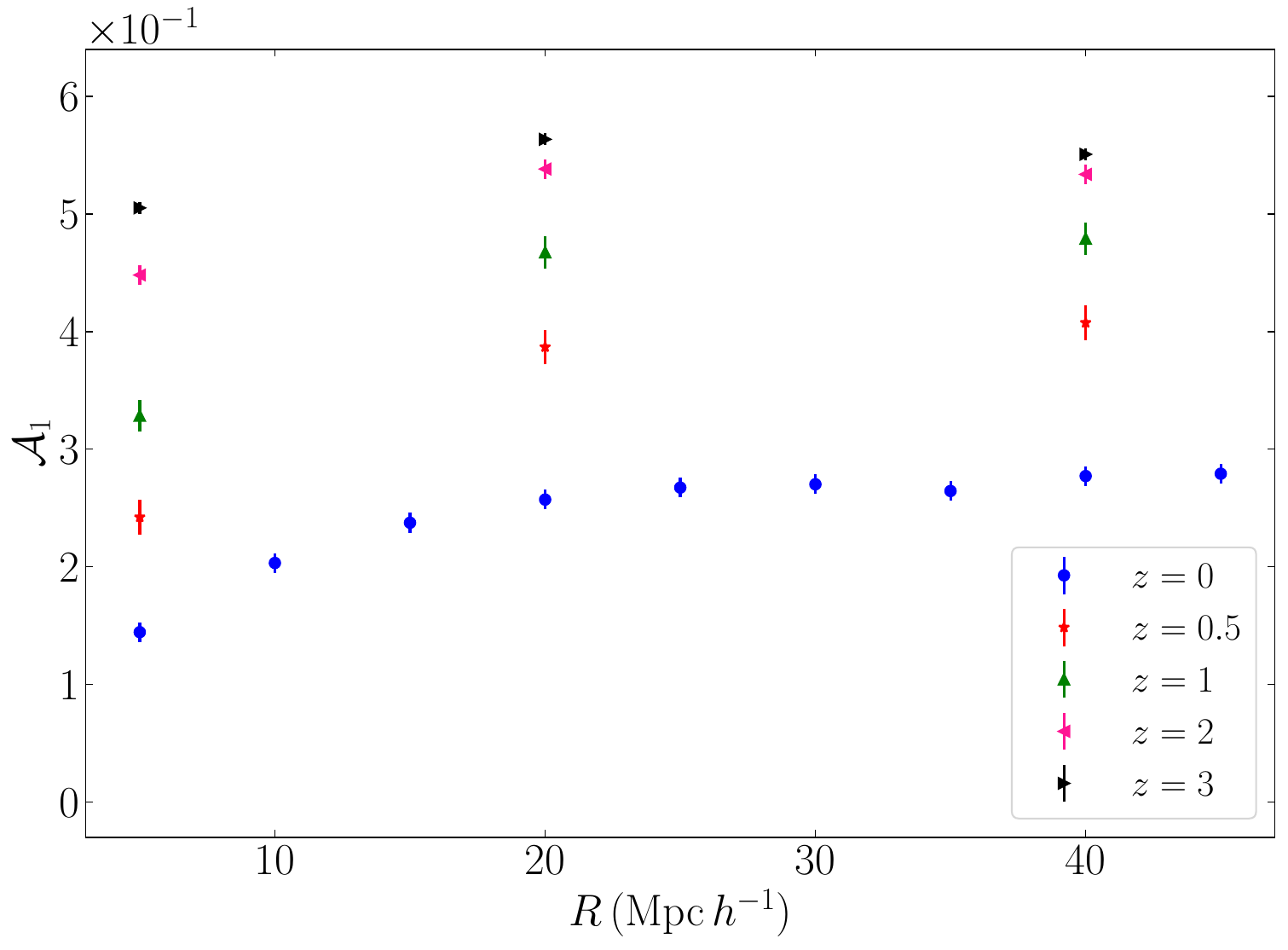}
            \includegraphics[scale=0.24]{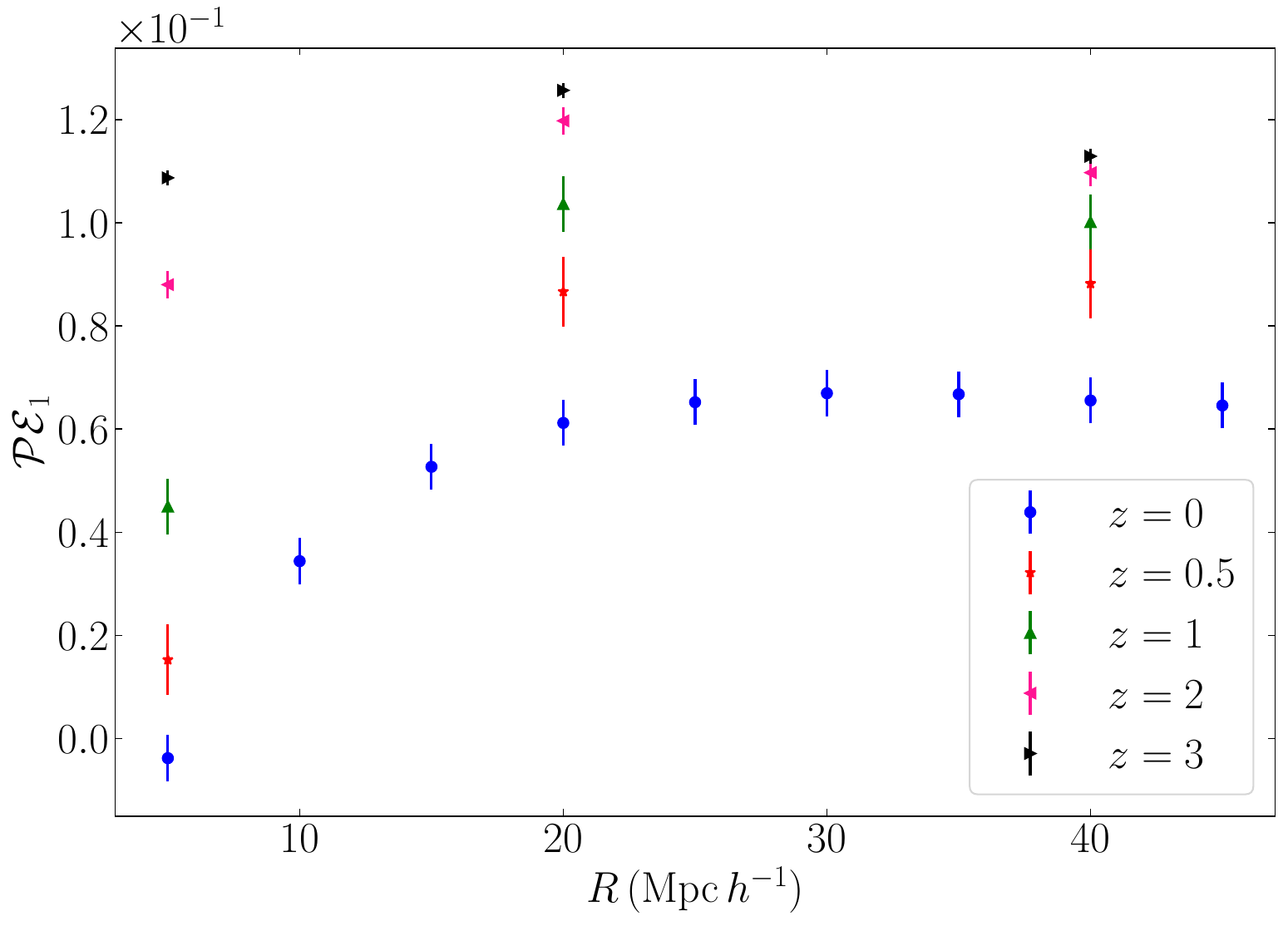}
            \includegraphics[scale=0.24]{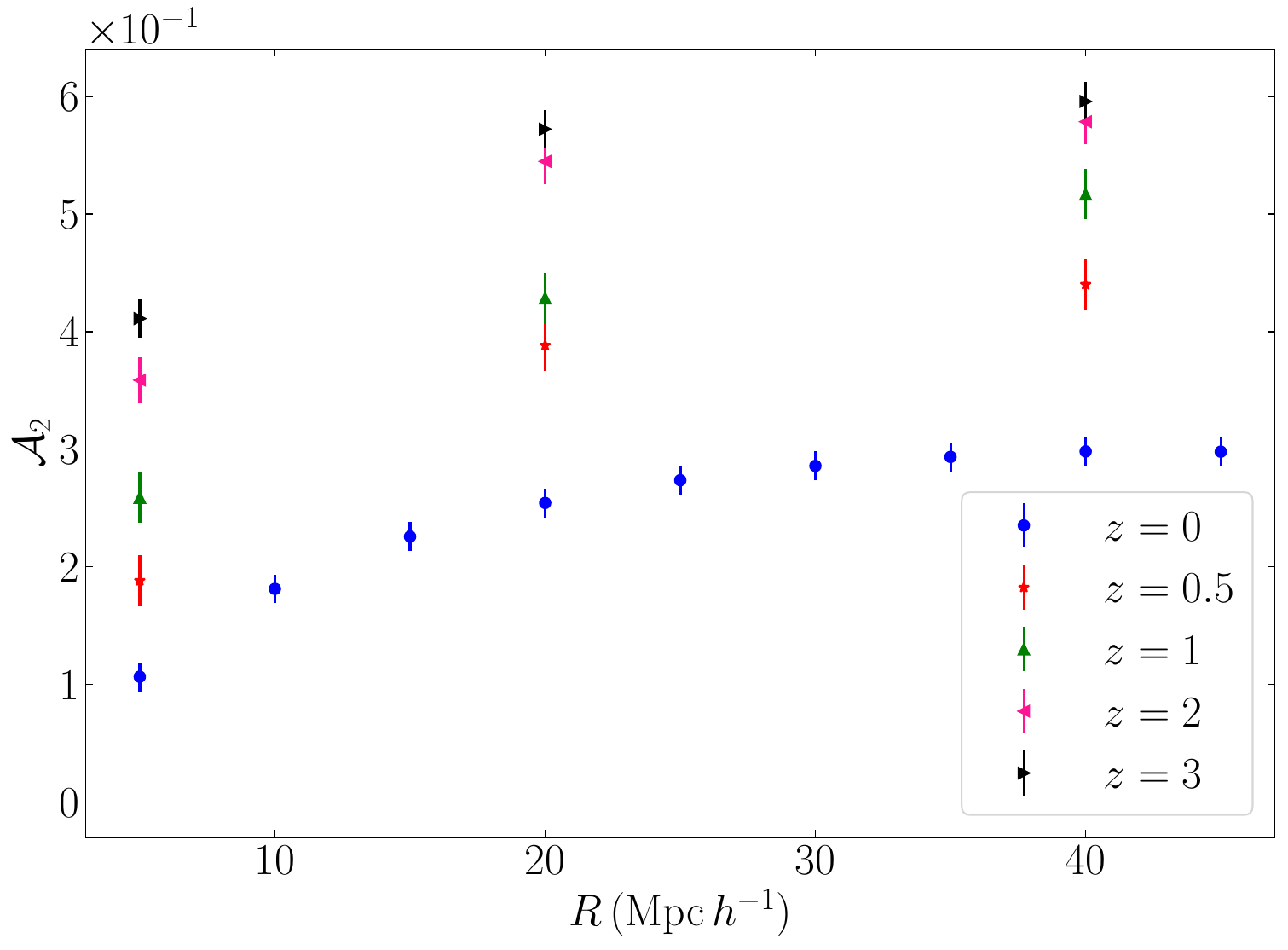}
            \includegraphics[scale=0.24]{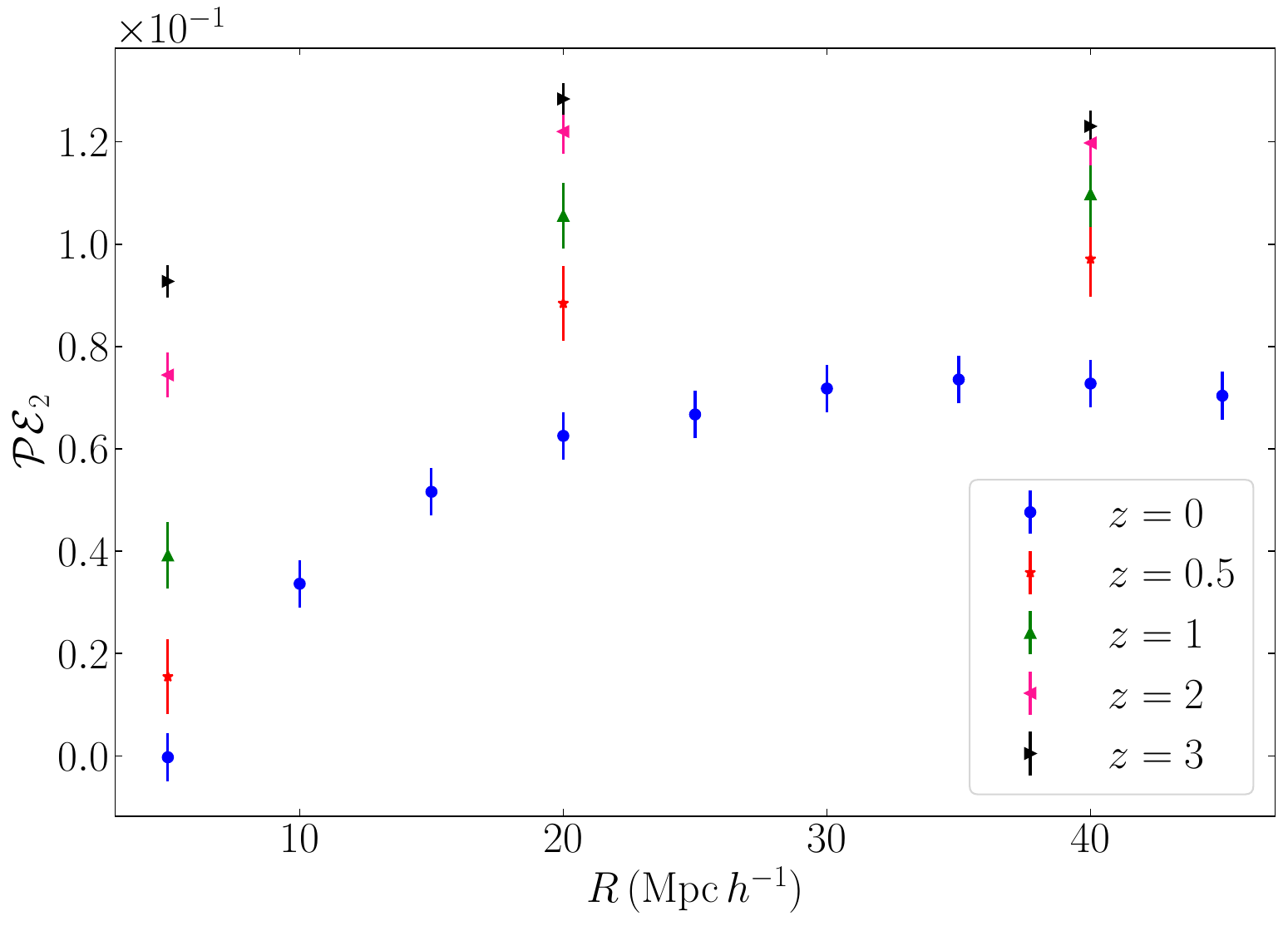}
        \end{center}
        \caption{The ratio of total the Betti number and the difference of the real and redshift spaces persistent entropy as a function of smoothing scale for five different redshift. The error-bar has been computed at $2\sigma$ level of confidence.}
        \label{fig:peA}
    \end{figure*}

    \begin{figure}
        \begin{center}
            \includegraphics[scale=0.3]{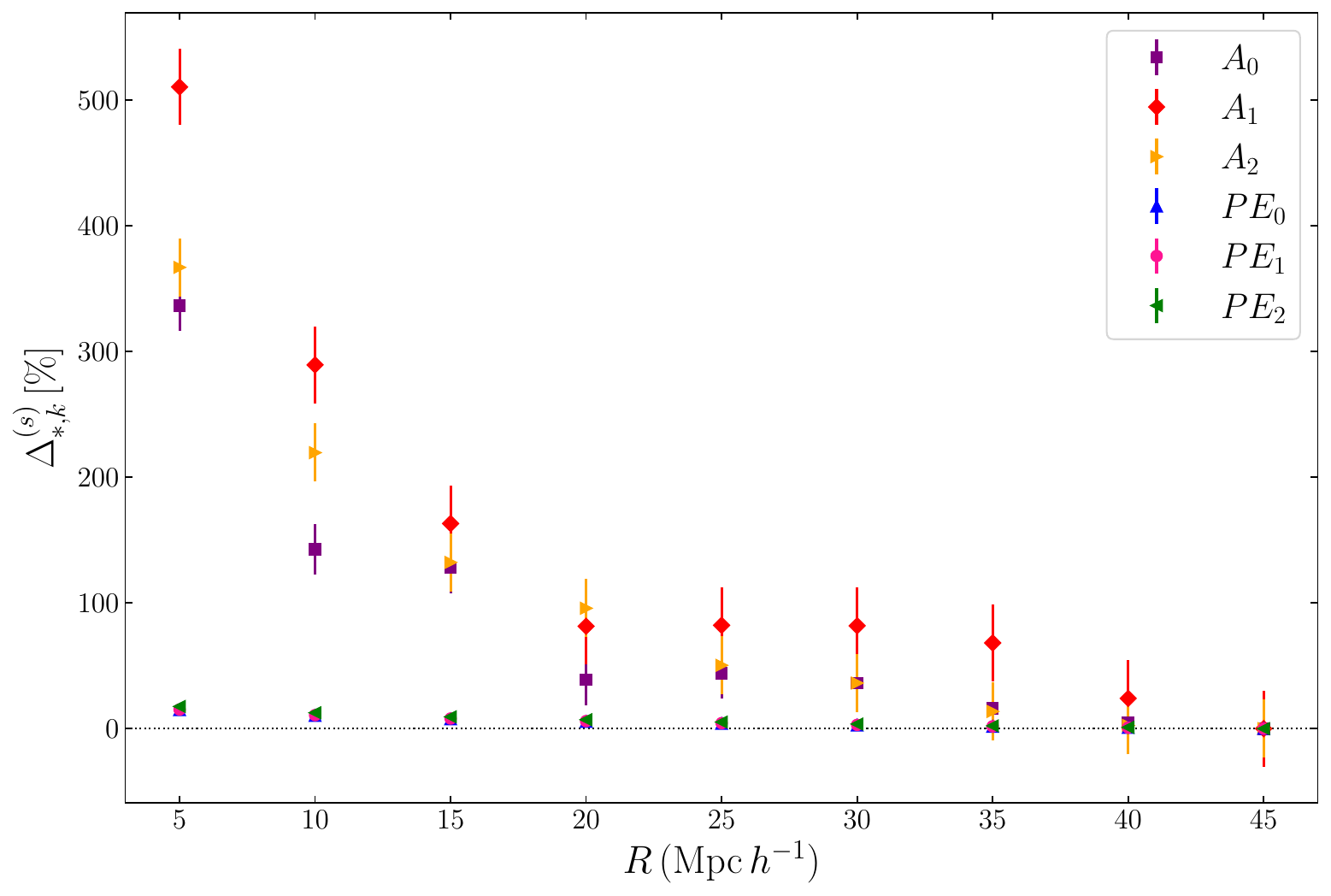}
            \includegraphics[scale=0.3]{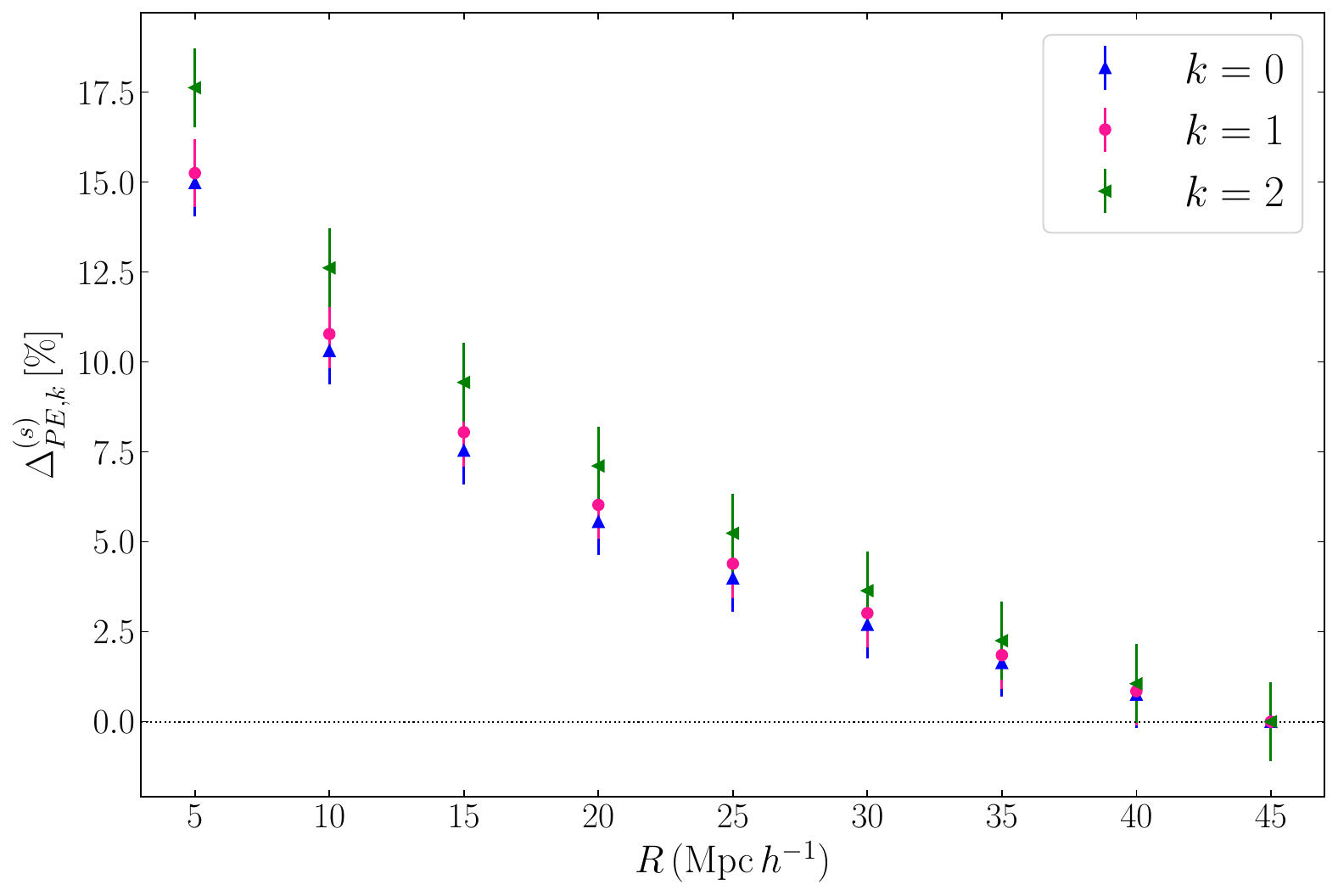}
        \end{center}
        \caption{The $\Delta^{(s)}_{A,k}(R)$ and $\Delta^{(s)}_{PE,k}(R)$ for the redshift space versus smoothing scale. Upper panel shows both $\Delta^{(s)}_{A,k}(R)$ and  $\Delta^{(s)}_{PE,k}(R)$ for better comparison. The maximum percentage of deviation for Persistent Entropy measure is negligible compared to $\Delta^{(s)}_{A,k}(R)$. To make more sense, the lower panel indicats the relative deviation of PE.The error-bar has been computed at $2\sigma$ level of confidence.}
        \label{fig:ratio}
    \end{figure}

    To explore the effects of RSD more quantitatively in terms of topological features and to identify a sensitive indicator (or one that is insensitive) to RSD, particularly its non-linear component, we compute additional complementary topological measures as defined in the section \ref{sec:ph}.  In Fig. \ref{fig:peA}, we depict the results for the difference between  persistent entropy of real and redshift space defined by $\mathcal{PE}_k \equiv PE_k^{(s)} - PE_k^{(r)}$, and the relative ratio of Area under the Betti number curve as: $\mathcal{A}_k \equiv \frac{A_k^{(s)}-A_k^{(r)}}{A_k^{(r)}}$. As previously stated, Redshift Space Distortion (RSD) is present across all scales. However, at small scales (wavenumber $\gtrsim 1$ Mpc$^{-1}\;h$), the Finger of God (FoG) effect is the dominant influence, whereas at larger scales, the linear Kaiser effect is essential. The modeling of the non-linear RSD effect is mathematically more complex than that of the Kaiser effect. Therefore, to determine the scale at which the non-linear RSD effect can be ignored and to identify which quantities are less sensitive to this phenomenon, we examine $\mathcal{PE}_k$ and $\mathcal{A}_k$ as a function of the smoothing scale across different redshifts.  The left and right column of Fig.~\ref{fig:peA} are devoted to $\mathcal{A}_k$ and  $\mathcal{PE}_k$, respectively.   For the matter density field simulated at $z=0$, one can obtain a trend for both introduced measures versus $R$. What we observe from this figure indicates that for nearly $R \gtrsim 30$ Mpc $h^{-1}$, the variations of $\mathcal{A}_k$ and $\mathcal{PE}_k$ as a function of $R$ begin to plateau. These results suggest that when viewed through the lens of topological features, only the linear Kaiser effect plays a significant role in the context of RSD for $R \gtrsim 30$ Mpc $h^{-1}$ at $z=0$, while the non-linear components are less influential. To formulate a more accurate assessment for higher redshifts, additional simulations are required, which have not been feasible for us due to a lack of computational resources. Nevertheless, it can be inferred that the dependence of $\mathcal{A}_k$ and $\mathcal{PE}_k$ on $R$ diminishes as the redshift increases (different symbols in Fig. \ref{fig:peA}). To ascertain the length scale at which non-linear effects manifest, we introduce two proper quantities as: $\Delta^{(s)}_{A,k}(R)\equiv \left[A^{(s)}_k(R)-A^{(s)}_k(R_{max}=45\;{\rm Mpc}\; h^{-1})\right]/A^{(s)}_k(R_{max}=45\;{\rm Mpc}\; h^{-1})$ and  $\Delta^{(s)}_{PE,k}(R)\equiv \left[PE^{(s)}_k(R)-PE^{(s)}_k(R_{max}=45\;{\rm Mpc}\; h^{-1})\right]/PE^{(s)}_k(R_{max}=45\;{\rm Mpc}\; h^{-1})$. To mitigate spurious results, such as those arising from size effects, the maximum smoothing scale utilized here is set to $R_{max} \lesssim (L_{box}/~20)$ Mpc $h^{-1}$. Fig. \ref{fig:ratio} shows that the total abundance of the reduce Betti number is significantly affected by non-linear RSD. On the other hand,  Persistent Entropy appears to be a relatively insensitive measure regarding non-linearity with the maximum percentage of deviation for $\Delta^{(s)}_{PE,k}$ is $\lesssim 20 \%$.

    \section{Summary and Conclusions}\label{sec:con}
    The qualitative and quantitative advancements in both existing and upcoming redshift surveys along with the progressive development of computational algorithms for  N-body simulations and statistical inferences, have inspired us to employ a topological-based data analysis framework. This approach was utilized to analyze the synthetic matter density field, which has been influenced by redshift space distortions. The central aim of this investigation was to analyze Persistent Homology and its complementary measures for the matter density field affected by RSD. More precisely, we were interested in determining the amount of variation in different topological measures due to the linear and non-linear components of redshift distortions. To achieve this goal, we constructed a dedicated pipeline including generating matter density fields by using Quijote N-body simulation in both real and redshift spaces  across various redshift epochs and smoothed by various values of smoothing scales; performing the super-level filtration and consequently making the excursion set for different thresholds; derivation of persistence diagram for connected regions, closed loops and closed surfaces; defining proper summary statistics according to assumed PH vectorization (Fig. \ref{fig:pipeline}). Our summary statistics was ${\rm PH}^{\rm{(s,r)}}:\left(\tilde{\beta}^{\rm{(s,r)}}_{k},\;  A^{\rm{(s,r)}}_k,\;  PE^{\rm{(s,r)}}_k   \right)$.

    We pursued  the impact of smoothing consequence and redshift evolution on the reduced Betti numbers in real space (Figs. \ref{fig:bcsreal} and \ref{fig:bcsrealz}).  The implementation of smoothing formalism has affected both the amplitude and the peak position of $\tilde{\beta}_k$.  Nevertheless, the maximum value of $\tilde{\beta}_k$ shows a notable degree of robustness against rising redshift, at least for the specific redshift range considered in this investigation. Among the various $\tilde{\beta}_k$ parameters in real space, the voids represented by $(\tilde{\beta}_2)$ exhibit a heightened sensitivity to changes in the $R$ and the redshift. Comparing the results for matter density field distorted by RSD with its counterpart in real space demonstrated that the Betti number curve was shifted to higher (lower) thresholds for $0$-hole (1- and 2-hole) (Fig. \ref{fig:realredshift}).  The closed surfaces (voids) were more sensitive to RSD for small smoothing scale, while the $\tilde{\beta}_0$ and $\tilde{\beta}_1$ were robust with respect to RSD, particularly for high smoothing scale (Fig. \ref{fig:realredshift}). 

    By definition $\mathcal{PE}_k$ and $\mathcal{A}_k$, we also quantified the RSD effect for different smoothing scales and redshifts (Fig. \ref{fig:peA}). Our results depicted that the linear Kaiser effect has a crucial role for $R \gtrsim 30$ Mpc $h^{-1}$ at $z=0$. Also the $R$-dependency of $\mathcal{A}_k$ and  $\mathcal{PE}_k$ decreases by increasing the redshift (different symbols in Fig. \ref{fig:peA}).
    To evaluate the length scale at which non-linear effects are present and to introduce a reliable measure that is not influenced by non-linearity, we define two so-called contrast quantities, $\Delta^{(s)}_{A,k}(R)$ and $\Delta^{(s)}_{PE,k}(R)$. Our analysis reveals that Persistent Entropy is an appropriate measure for probing the linear Kaiser effect, with the maximum deviation percentage for $\Delta^{(s)}_{PE,k}$ is $\lesssim 20 \%$ (Fig. \ref{fig:ratio}).

    The directions for further research on this topic present some promising approaches of increasing model accuracy, extending the domain of applicability, and utilizing advanced methodologies.Specifically, the investigation of anisotropy and non-Gaussianity of LSS due to RSD are valuable topics as the complementary parts. This can be accomplished by designing appropriate measures through PH that are sensitive to directional behaviors and non-Gaussianity. A feasible idea is to employ pre-processing techniques, such as curvelet analysis and various directional filtering approaches, on the constructed map in redshift space for different directions. The outputs generated from this procedure can be used an input to compute the PH feature vector.  Any deviation obtained for topological properties for different directions may be potentially assumed as the directional indicator measures; Going beyond the plane-parallel approximation and accounting the spectroscopic and photometric errors; Fisher Forecast for Parameter Constraints is also useful for refining  parameter constraints (particularly constraining on $\mathcal{B}$ (see \cite{Appleby2023,appleby2019ensemble,appleby2018minkowski,2024ApJ...963...31K})); enhancing model accuracy and reducing uncertainty through more precise parameter estimations; Implementing Bayesian Model Averaging technique  will address model uncertainty by averaging over multiple models. This approach attempts  to improve predictive performance and mitigate model selection bias; Simulation-Based Inference method such as Approximate Bayesian Computation or Markov Chain Monte Carlo enables us to adopt Likelihood-free inference and model fit through robust simulations  \cite{tejero2020sbi,2016arXiv160506376P,2019MNRAS.488.4440A,2020PNAS..11730055C}; In this study, we assumed the matter density field, while other tracers such as galaxies has motivation to assess from our approach. In this case, the influence of bias on the PH vectorization for  different tracers could be incorporated.

\begin{acknowledgements}
We  thank the Quijote team for sharing its simulated data sets and providing extensive instruction on how to utilize the data. The PH calculations were performed by using \texttt{Cubical ripser} \cite{2020arXiv200512692K}, and  \texttt{GUDHI} \cite{maria2014gudhi}  python packages.

\end{acknowledgements}

\appendix

\section{From data to Topological Data Analysis}
Various kinds of data are collected  through experimental setups and surveys, furthermore our aim is to develop  pipelines for information extraction as effectively as possible. To achieve this, understanding the different types of data from a mathematical  perspective allows us to develop appropriate mappings between them, making them suitable for further analysis.  In particular, we are able to assign distinct shapes to the data within the context of topological data analysis.

Generally, one can classify the data type into {\it time series}, {\it field}, {\it point cloud} and {\it network}. A multi-variable time series in $(D+1)-$dimension can be expressed by \cite{kantz2003nonlinear}:
\begin{equation}
\mathcal{X}=\left\{\vec{x}^{(j)}\;|\; \vec{x}^{(j)}=\left(x^{(j)}(t_i)\right)_{i=1}^{\mathcal{T}}\right\}_{j=1}^D\equiv\left\{x^{(j)}(t_i)   \right\}_{(i=1,j=1)}^{(\mathcal{T},D)}
\end{equation}
A multi-variable field in $(D+N)-$dimension is given by \cite{adler81}:
\begin{equation}
\mathfrak{F}=\left\{\mathcal{F}_j\;|\; \mathcal{F}_j:\Pi_j\to\mathbb{R}, \;\Pi_j\subset\mathbb{R}^{N} \right\}_{j=1}^{D}
\end{equation}
From mathematical perspective, a generic point cloud in $D-$dimension with size $N$ is represented by:
\begin{equation}
\mathbb{X}=\left\{x_i\;|\; x_i\equiv\left(x_i^{(j)}\right)_{j=1}^{D}, \;x_i^{(j)}\in \mathbb R\right\}_{i=1}^{N}
\end{equation}
A network is denoted by $G=(V,E,w)$ in which $V\equiv\left\{v_i\right\}_{i=1}^N$ is the vertex (node) set, $E=V\times V$ is edge (link) set and weight function is defined as $w:\; E\to \mathbb{R}$. A variety of methods have been developed to establish connections among different types of data. The application of recurrent plots represents a practical method for the development of a field from time series data \cite{marwan2008historical}. Various mapping techniques, such as time delay embedding \cite{takens2006detecting} and state space, facilitate the conversion of time series into point clouds \cite[and references therein]{2023PhRvE.107c4303M}. To construct a network from these time series, one can utilize the visibility graph \cite{lacasa2008time} and correlation network \cite{zou2019complex}. The excursion set is instrumental in creating point clouds from the field \cite{adler81}. Additionally, recurrent networks are employed to form networks from the generated point clouds \cite{chen2018recurrence}. Prior to topological based data analysis, a proper tessellation approach is utilized, based on the constructed sets that include field, network, and point cloud.

\section{Homology Group in nutshell}

Unlike geometrical invariants that remain unchanged under congruence, we claim that two spaces are topologically equivalent if and only if their topological characteristics (global features) are preserved under homeomorphisms. Homology theory plays an important and essential role in the mathematical characterization of the fundamental components of a typical topological space, as it elucidates the connectedness of the underlying structure. For the purpose of determining the Homology group of the underlying topological space, one must first clarify the essential building blocks of algebraic topology. For the pedagogical purpose, we first explain the $k$th Betti number ($\beta_k(\mathcal M)$) of a simplicial complex ($\mathcal M$) as the dimension of $k$-Homology group:
\begin{equation}
\beta_k(\mathcal M) \equiv{\rm dim}[\mathcal{H}_k(\mathcal{M})]
\end{equation}
The quotient group of the $k$-cycles group $(\mathcal{Z}_k(\mathcal{M}))$ by the $k$-boundary group $(\mathbb{B}_k(\mathcal{M}))$ represents the  $k$-Homology group as:
\begin{equation}
\mathcal{H}(\mathcal{M})\equiv \mathcal{Z}_k(\mathcal{M})/\mathbb{B}_k(\mathcal{M})
\end{equation}
The $k$-cycle group is also defined as the collection of all $k$-cycles as:
\begin{equation}
\mathcal{Z}_k(\mathcal{M})\equiv\{c_k \in C_k(\mathcal{M})|\partial_k(c_k)=\emptyset\}
\end{equation}
in such that the $k$-chain group is defined by:
\begin{equation}
C_k(\mathcal{M})\equiv\left\{c_k|c_k=\sum_{i=1}^{|\Sigma_k(\mathcal{M})|} a_i\sigma_k(i) \right\}
\end{equation}
in this context, $\sigma_k(i)$ represents the convex hull formed by $k+1$ independent points in $\mathbb{R}^D$ known as $k$-simplex. The mathematical form is written as $\sigma_k\equiv[v_0,v_1,...,v_k]\subseteq \mathbb{R}^D$. Consequently, a 0-simplex corresponds to a single point, a 1-simplex is defined as a line segment, a 2-simplex refers to a filled triangle, a 3-simplex denotes a filled tetrahedron, and this pattern continues for higher dimensions. An alternative form of structures that facilitate the tessellation of topological spaces is called cubical simplex. In this approach the multiple-dimensional cubes are used to construct these structures. For instance, $0-$dimensional cube is represented by point, $1-$dimensional cube is line segment, square is 2-dimensional cube, and 3-dimensional cube is indicated simply by cube and so on. A $k$-simplex's $\ell$-face is also defined as a subset of $\ell + 1$ vertices selected from the $k$-simplex's vertices while satisfying the condition $0 \leq \ell \leq k$. We can now characterize the simplicial complex, denoted as $\mathcal{M}$, by the following principle: every $\ell$-face of a $k$-simplex is included in the complex, and the intersection of any two simplices, $\sigma_k$ and $\sigma_m$, is a non-empty set that constitutes an $\ell$-face of both simplices. The $k$-ordered sub-collection of complexes is represented  by $\sum_k(\mathcal{M})\equiv\{\sigma \in \mathcal{M}\;|\;{\rm dim}(\sigma)=k\}$. Therefore  $k$-simplex is $\sigma_k(i)\in\Sigma_k(\mathcal{M})$. The linear combination of $k$-simplices of $\mathcal{M}$ defines the $k$-chain as $c_k \equiv\sum_{i=1}^{|\sum_k(\mathcal{M})|} a_i\sigma_k(i)$ and $a_i\in \mathbb{Z}_2$ such that $\mathbb{Z}_2\equiv[0,1]$. The boundary operator ($\partial_k$)  maps the $\sigma_k$ to its boundary according to $\partial_k(\sigma_k)\equiv
\sum_{j=0}^{k}(-1)^j [v_0,v_1,...,v_{j-1},v_{j+1},...,v_k] \subseteq \sigma_k$. Subsequently, the $k$-boundary group, $\mathbb{B}_k(\mathcal{M})$, is a collection of $k$-boundary denoted by $b_k$ set as:
\begin{equation}
\mathbb{B}_k(\mathcal{M})\equiv\{b_k(i)\}_i^{|\mathbb{B}_k(\mathcal{M})|}\subseteq C_k(\mathcal{M})
\end{equation}
In accordance with the previous statement, the PH explores the generation ($birth$) and dissolution ($death$) of constructed topological invariants associated with homology classes within the context of a mathematical process identified as filtration. Various PH vectorizations have been introduced based on the specific objectives of the study \cite{adams2017persistence,2021arXiv210908721P,2023arXiv231113520J,2024arXiv240313985Y}.

In summary, it is essential to recognize that many computational algorithms have been created to construct complexes from typical data sets, enabling the examination of associated Persistent Homology. The following enumeration includes several prominent algorithms and methods: Vietoris-Rips Complex, \u{C}ech Complex, Alpha Complex, Sparse Rips Complex, Delaunay Complex, Witness Complex, Greedy Witness Complex, and the Mapper Algorithm \cite{zomorodian2004computing,zomorodian2005topology11,edelsbrunner2008persistent,adler2010persistent,edelsbrunner2022computational,dey2022computational}.


\end{document}